\documentclass[onecolumn,authoryear]{els-mrw}

\usepackage{graphicx} % Required for inserting images

\usepackage{url}
\usepackage{comment}
\usepackage{color}

\usepackage{hyperref}
\hypersetup{
    colorlinks = true,
    citecolor  = blue,
    linkcolor  = blue,
    urlcolor  = blue
}

\usepackage{physics, mathtools, braket}
\usepackage{amsmath,amsfonts,amssymb,amsthm}

\newcommand{\Sch}{ Schr\"{o}dinger }

\newcommand{\NpartNhole}[1]{{#1}p-{#1}h}%
\newcommand{\NholeNpart}[1]{{#1}h-{#1}p}%

\usepackage[most]{tcolorbox}
\usepackage{xcolor}

\newtcolorbox{keypoint}{
  colback=gray!15,   % grey background
  colframe=gray!60,  % grey border
  boxrule=0pt,       % no visible frame
  left=6pt,
  right=6pt,
  top=6pt,
  bottom=6pt
}

\title{EncyclopediaNP}

\begin{document}

%\chapter{Theoretical  description of giant resonances in nuclei}

%\chapter{Giant resonances in nuclei from first principles}

%\chapter{Giant resonances and their theoretical description from first principles} 
\chapter{Nuclear giant resonances from first principles} 

\author[1,2]{Sonia Bacca}
\author[1]{Francesco Marino}
\author[3,4]{Andrea Porro}

\address[1]{\orgdiv{Institut f\"{u}r Kernphysik and PRISMA+ Cluster of Excellence}, \orgname{Johannes Gutenberg-Universit\"{a}t Mainz}, \orgaddress{55128 Mainz, Germany} }
\address[2]{\orgdiv{Helmholtz-Institut Mainz}, \orgname{Johannes Gutenberg-Universit\"{a}t Mainz}, \orgaddress{D-55099 Mainz, Germany}}
\address[3]{\orgname{Technische Universit\"at Darmstadt}, \orgdiv{Department of Physics}, \orgaddress{64289 Darmstadt, Germany}}
\address[4]{\orgname{ExtreMe Matter Institute EMMI}, \orgdiv{GSI Helmholtzzentrum f\"ur Schwerionenforschung GmbH}, \orgaddress{64291 Darmstadt, Germany}}

\articletag{Chapter Article tagline: update of previous edition, reprint..}

\maketitle

\section*{Glossary}
\noindent{\bf Nucleons:} Particles of semi-integer spin which compose the nucleus. The term is used to denote protons and neutrons together. \\
\noindent{\bf Collective excitations:} Excitation modes characterized by the coherent participation of many particles. In nuclear systems, the relevant particles are the nucleons.\\
\noindent{\bf Giant dipole resonances:} Excitation modes in which all protons oscillate in phase against all neutrons.\\
\noindent{\bf Giant monopole resonances:}
Excitation mode that leads to a uniform compression and expansion of the entire nucleus as a ``breathing mode".

\section*{Acronyms}
\begin{tabular}{ll}
$\chi${\bf EFT}& Chiral effective field theory \\
{\bf CC}& Coupled-cluster\\
{\bf CCSD}& Coupled-cluster singles and doubles\\
{\bf CCSDT-1}& Coupled-cluster singles and doubles with leading order triples\\
{\bf EDF}& Energy-density functional\\
{\bf EOM}& Equation-of-motion\\
{\bf GCM} & Generator coordinate method\\
{\bf GDR} & Giant dipole resonance\\
{\bf GMR} & Giant monopole resonance\\
{\bf HF}& Hartree-Fock\\
{\bf HFB}&Hartree-Fock-Bogoliubov\\
{\bf HF-RPA} & Hartree-Fock random phase approximation\\
%{\bf HF-SRPA} & Hartree-Fock second random phase approximation\\
{\bf IMSRG} & In-medium similarity renormalization group\\
%{\bf IM-RPA} & In-medium  random phase approximation\\
{\bf LIT} & Lorentz integral transform\\
{\bf LIT-CC} & Lorentz integral transform coupled-cluster\\
{\bf NCSM} & No-core shell model\\
{\bf PGCM} & Projected Generator Coordinate Method\\
{\bf QRPA} & Quasi-particle random phase approximation\\
{\bf RPA} & Random phase approximation\\
{\bf SCGF} & Self-consistent Green's functions\\
{\bf SRG} &  Similarity renormalization group
\end{tabular}

\begin{abstract}
This chapter presents an ab initio perspective on giant resonances in atomic nuclei and surveys the principal theoretical frameworks that aim to describe these collective excitations from first principles. While the study of nuclear giant resonances has traditionally been dominated by the energy density functional approach, recent years have witnessed the development of advanced many-body approaches grounded directly in realistic nuclear interactions, namely, Hamiltonians that reproduce nucleon–nucleon phase shifts and accurately describe the binding energies of light nuclei.
Within this modern framework, we review the main many-body methods currently used to compute nuclear response functions.
These include the random phase approximation, the Lorentz integral transform coupled-cluster theory, the projected generator-coordinate method, and the self-consistent Green’s functions approach.
After giving a general conceptual and historical overview of giant-resonance phenomena, we outline the theoretical foundations and computational implementations of each method.
We conclude with a critical comparison of their predictions for selected benchmark nuclei, $^{16}$O and $^{40}$Ca,  emphasizing points of agreement and divergence, while maintaining  a close connection to the relevant experimental observables.
\end{abstract}

%\section*{Key points}

\begin{keypoint}
\textbf{Key Points:}\\ 
\begin{itemize}
\item  Ab initio methods  aim to calculate nuclear properties starting directly from realistic interactions among nucleons, without introducing parameters adjusted to specific nuclei. 
\item The comparison of theoretical calculations with experimental data provides stringent tests on our understanding of the strong force.

\item Giant resonances in nuclei are described through the concept of response functions, which can be in first place derived using linear response theory and calculated within a variety of many-body frameworks.

\item Ab initio calculations of giant resonances in nuclei have shown that collectivity emerges from microscopic nuclear forces.

\item While enormous progress has been made over the past decade in the ab initio theoretical description of giant resonances, further work is required to strengthen the agreement among different methods and to thoroughly assess theoretical uncertainties.
\end{itemize}
\end{keypoint}

\section{Introduction} 

Giant resonances are among the most striking collective phenomena that occur in the atomic nucleus. They can be thought of as coherent oscillations in which many, or even all, nucleons participate. In such a mode, the nucleus as a whole is excited in a correlated way, giving rise to strong, broad peaks in the excitation spectrum. Because of this collectivity, giant resonances are often compared to the vibrations of a macroscopic object, for instance the oscillation of a drop of liquid after it is disturbed. The analogy is helpful: in both cases, the motion of the microscopic constituents is strongly correlated, producing a large-amplitude response at well-defined frequencies.
From the characteristic frequencies of these resonances, nuclear physicists seek to infer key properties of the nuclear interior, such as the effective interactions among nucleons and the bulk response of nuclear matter.

Experimentally, giant resonances were first identified in the  1940s, when early photoabsorption measurements revealed a broad resonance structure located at around 10-20 MeV of excitation energy. 
First observed in uranium by \cite{BaK47},
%Baldwin and Klaiber in 1947
this prominent structure has subsequently been seen in nuclei across the entire nuclear chart.
This feature was later recognized as the giant dipole resonance, in which protons and neutrons oscillate against each other.
The discovery was remarkable: instead of being distributed over many small transitions, the dipole strength appeared concentrated in a single broad structure  that carried most of the expected strength. This observation established giant resonances as a new class of nuclear excitations and stimulated decades of experimental and theoretical work.

Following the discovery of the giant dipole resonance, other modes were identified. The giant monopole resonance, often called the ``breathing mode," corresponds to a uniform compression and expansion of the entire nucleus as explained by \cite{Youngblood1977}. The giant quadrupole resonance, as described by  \cite{vanderWoude1987}, involves collective oscillations with a quadrupole shape, and higher multipolarities have also been observed. Moreover, spin–isospin giant resonances, such as the Gamow–Teller resonance explained (\cite{Osterfeld1992}), have played a central role in both nuclear physics and astrophysics. 
Collective modes provide a powerful window into bulk nuclear properties, such as the incompressibility and the symmetry energy of nuclear matter, which can be correlated with the properties of the monopole (\cite{Blaizot1980,Garg18a}) and dipole resonances (\cite{rocamaza2018}), respectively.

Theoretically, giant resonances occupy a special place because they connect single-particle dynamics with collective behavior. Already in the 1940s and 1950s, simple models were proposed to explain the giant dipole resonance (GDR). Two classic pictures stand out: the \cite{GoT48} model, which describes the resonance as an out-of-phase oscillation of the proton and neutron fluids, and the \cite{steinwedel1950} model, which pictures protons and neutrons oscillating within a nuclear volume with a standing-wave character. 
Although schematic, these models captured essential features of the resonance and highlighted its collective nature. 
Since its discovery, the giant dipole resonance  has been the subject of numerous theoretical investigations aimed at explaining both its centroid and width within collective and microscopic many-body frameworks. 
For example, its width was modeled from the transfer of energy from a vibration to other modes of nuclear motion by (\cite{Speth1981, Bertsch1983}). 
Self-consistent mean-field theories have been widely employed to describe the GDR, see, e.g., (\cite{Paar:2007bk,Erler2011, Nakatsukasa2012}). Additionally, the effects of the continuum, first explored in (\cite{Shlomo}), and of particle-phonon coupling, see, e.g., (\cite{Colo2001,LyT12}), have been extensively studied.
These approaches have been very successful in accounting for global properties of giant resonances, but they typically rely on phenomenological interactions or energy functionals, which contain parameters tuned to experimental properties of finite (typically, mid-mass or heavy) nuclei.
The interested reader can find a recent review on modern applications of the random phase approximation (RPA) based on energy density functional (EDF) theory in (\cite{Schunck19a,Colo2022Handbook}).
Giant resonances are discussed from an experimental perspective in (\cite{HarakehvanDerWoude,Bracco2019,vonneumanncosel2025electricdipolepolarizabilityconstraints}).

Over the last decade, the field of nuclear theory has advanced to the point of an ab initio description of giant resonances—an achievement that represents a substantial paradigm shift.
%In recent years, the field has advanced to the point of enabling an ab initio description of giant resonances—an achievement that represents a substantial paradigm shift. 
Reporting on this subject in a pedagogical way is the focus of this review. The Latin term ab initio, meaning “from the beginning” or “from first principles” refers to approaches that aim to calculate nuclear properties starting directly from realistic interactions among nucleons, without introducing parameters adjusted to specific nuclei. 
The guiding principles and overall spirit of the ab initio framework are clearly articulated in the recent work of (\cite{Hergert2020,Ekstrom2023,papenbrock2024}) and in ~(\cite{ChapterHergert}).
%\textcolor{red}{Hergert's contribution to this Encyclopedia}.  
In the ab initio approach, nuclear forces are typically derived consistently from the underlying theory of the strong interaction, quantum chromodynamics (QCD), often through effective field theories such as chiral effective field theory ($\chi$EFT), see e.g. (\cite{MACHLEIDT2024104117,Epelbaum2024}). 
Over the past few decades, $\chi$EFT has established itself as a powerful tool to link  properties of nucleons to QCD in a systematically improvable manner.
Nuclear forces derived from $\chi$EFT  are then used in many-body methods that attempt to solve the nuclear Schrödinger equation with controlled and systematically improvable approximations.

Ab initio many-body methods start from the Schr\"odinger equation 
\begin{equation}
\label{eq:Lipp}
 H | \Psi_{\nu} \rangle = E_{\nu} | \Psi_\nu \rangle \,
\end{equation}
of $A$ nucleons  ($Z$ protons and $N$ neutrons) interacting with each other through a Hamiltonian $H$, with the goal of finding the energies $E_\nu$ and the wave functions $|\Psi_\nu \rangle$, for the ground state as well as for excited states.
The nuclear Hamiltonian $H$ is given by
\begin{equation}
\label{eq:hamiltonian}
 H = T_K+ \sum_{i<j}^A V_{ij} + \sum_{i<j<k}^A W_{ijk} + \dots \ ,
\end{equation}
where $T_K$ is the intrinsic kinetic energy, while $V_{ij}$ and $W_{ijk}$ are the nucleon-nucleon and three-nucleon forces, respectively, which are systematically derived in $\chi$EFT. Higher order operators represented by the dots in Eq.~(\ref{eq:hamiltonian}) may exist in the Hamiltonian, but are most often neglected.

In $\chi$EFT, nuclear interactions are derived in a low-momentum expansion, where the QCD Lagrangian is mapped to an effective theory formulated in terms of the emergent
degrees of freedom relevant for nuclear physics, that is, nucleons and pions.
A hierarchy of contributions to the nuclear Hamiltonian is found in powers of $Q/\Lambda$, where $Q$ denotes a characteristic scale of nuclear systems and $\Lambda$ represents the breakdown scale of the effective field theory.
%organized as an expansion in powers of \(Q/\Lambda\), where \(Q\) denotes a characteristic low--momentum scale of nuclear systems and \(\Lambda\) represents the breakdown scale of the effective field theory. 
The various contributions entering the two-- and three--nucleon interactions, \(V_{ij}\) and \(W_{ijk}\), are represented through  corresponding Feynman diagrams. Short-range physics that cannot be resolved explicitly is absorbed into low-energy constants, whose values are typically determined by fits to selected experimental observables, including nucleon-nucleon phase shifts and possibly the binding energy of light nuclei. Multiple strategies exist for optimizing and constraining these constants, for which we refer the reader to ~(\cite{ChapterHebeler}).
%\textcolor{red}{Hebeler and Ekstr\"om's contribution to this Encyclopedia}. 
In this work, we will just present results based on a specific choice of a few \(\chi\)EFT-inspired Hamiltonians.
However, we remark that in the ab initio framework the achievable precision in 
the computation of a given observable is set partly by the truncation and structure of the underlying \(\chi\)EFT and partly by the accuracy with which the few-- or many--body problem can be solved.
Here, we will not analyze the former, and only partially discuss the latter. A full, comprehensive uncertainty quantification for ab initio computations of giant resonances in nuclei is still missing.

In this Encyclopedia Chapter, we will focus on the ab initio description of giant resonances in nuclei. The motivation for studying giant resonances with ab initio methods is twofold. First, such studies connect collective nuclear behavior—traditionally described with macroscopic or phenomenological models—directly to the underlying microscopic interactions. If one can compute the spectrum of collective resonances starting only from two- and three-nucleon forces constrained by few-body data, this provides a stringent test of our understanding of nuclear forces. Second, giant resonances are sensitive to bulk nuclear properties, such as the incompressibility of nuclear matter and the density dependence of the symmetry energy, which are crucial ingredients in modeling astrophysical objects such as neutron stars (\cite{rocamaza2018,Garg18a,Lattimer21a}). Hence, giant resonances provide a way of connecting laboratory-scale nuclear observables to the macroscopic scale of astrophysics.
From a practical perspective, giant resonances pose a formidable challenge for ab initio approaches. They lie at high excitation energies, often in the continuum, and involve strong correlations among many nucleons. Describing them requires both sophisticated theoretical methods and substantial computational resources. Nevertheless, advances in many-body techniques—such as coupled-cluster theory (\cite{Miorelli2018}), the in-medium similarity renormalization group (\cite{Parzuchowski16a}), the projected generator coordinate method (\cite{Porro24a}), and self-consistent Green’s functions theory (\cite{Raimondi2019})—together with modern high-performance computing, have opened the door to tackling these collective excitations in medium-mass nuclei and beyond.

In summary, giant resonances are collective oscillations of the nucleus that have fascinated nuclear physicists since their discovery. They offer deep insights into nuclear structure and the fundamental properties of nuclear matter. Historically, they have been understood through simple models and phenomenological approaches, which still play an important role in developing physical intuition. Today, however, the frontier lies in describing them within the ab initio paradigm, where collective excitations emerge naturally from microscopic nuclear forces and the correlations they induce among nucleons. This effort is not just a refinement of previous work; it represents a major step toward connecting the emergent collective behavior of the nucleus with the fundamental theory of the strong interaction.

\section{General principles}
Response functions, also called strength functions, are fundamental tools in nuclear physics and a key concept for the theoretical description of giant resonances.
In this Section, we first introduce response functions (Sec.~\ref{sec: definition response}) and motivate their definition by means of the linear response theory framework (Sec.~\ref{sec: Linear response theory}).
We further elucidate the connection between strength functions and experimental observables (Sec.~\ref{sec: connection exp}) by illustrating, in a representative case, how inelastic cross sections—the quantities directly measured—are naturally linked to the underlying response functions.

\subsection{Response function}
\label{sec: definition response}

The response function of a nucleus (described by a Hamiltonian $H$) with ground-state wave function $\ket{\Psi_0}$ and ground-state angular momentum $J_0$ is defined as 
\begin{equation}
\label{eq: def response function}
  R(\omega)=\frac{1}{2J_0+1}
  \sum_{\nu>0} 
  |\langle \Psi_0 | O |\Psi_\nu\rangle|^2 
\delta (E_\nu -E_0 -\omega)\,,
\end{equation}
where $O$ is the excitation operator with which the nucleus is probed, $\omega$ is the excitation energy, and $\Psi_\nu$ denotes the nuclear excited states with energies $E_\nu$.
Note that the sum $\nu>0$ extends over the whole spectrum of excited states of the nucleus, including both bound and continuum states.
The fact that continuum states are included can be easily understood as giant resonances are excitations lying above the particle-emission threshold, where a proton or a neutron is kicked out of the nucleus. 
Continuum states corresponding to all possible partitions of the nucleus into different fragments  should be, in principle, included.
However, explicit calculations of such channels are only feasible for up to five particles, see, e.g., Ref. (\cite{LAZAUSKAS2019335}), and approximations are thus required to deal with the problem, as we discuss in Sec.~\ref{sec: Many-body methods}.

Response functions characterize the probability distribution of a given transition and are essential not only for understanding nuclear structure and reaction mechanisms, as explained by \cite{BaccaPastore2014}, but also for applications in diverse areas such as nuclear astrophysics (\cite{goriely23}), medical isotopes production (\cite{wang22}), fission (\cite{chadwick11}), fusion reactor technologies (\cite{meschini23}), and nuclear waste transmutation (\cite{salvatore11}).

\subsection{Linear response theory}
\label{sec: Linear response theory}
The internal properties of a quantum-mechanical system can be accessed by acting on it with an external probe.
Provided that the coupling between the nucleus (or, in general, the many-particle system) and the probe can be considered ``weak", the process can be modeled by treating the external field as a perturbation and exploiting the framework of linear response theory.

The picture of response theory relies on separating the unperturbed system, described by a Hamiltonian $H$, from the external field, which is modeled by an operator $\mathcal{V}(t)$. 
Let us assume that the system lies initially in its unperturbed ground state $\ket{\Psi_0}$, when, at time $t=0$, the external field $\mathcal{V}(t)$ is switched on. A simple example is that of a target nucleus initially at rest, which at a well-defined initial time, say $t=0$, is struck by an incoming beam.
At $t>0$, the dynamics of the system is then governed by the time-dependent Hamiltonian $H(t) = H + \mathcal{V}(t)$.
For convenience, we follow (\cite{giuliani_vignale_2005}) and consider, without loss of generality, an external perturbation $\mathcal{V}(t) = v(t) B$, separating the quantum-mechanical operator $B$ from the dependence on time and on the magnitude of the field $v(t)$. 
Also, we consider only one-body perturbations for simplicity.

Our goal is to study the effect of the perturbation $\mathcal{V}(t)$ on some measurable quantity, which we describe by the expectation value of a properly chosen operator $A$ on the time-evolved state $\ket{\Psi(t)}$.
In particular, the perturbing field induces a fluctuation $\delta A(t)$ in the expectation value of $A$, 
\begin{align}
    \label{eq: def delta Ot}
    \delta A(t) = \mel{\Psi(t)}{A}{\Psi(t)} - \mel{\Psi_0}{A}{\Psi_0},
\end{align}
where $\ket{\Psi(t)}$ is determined by the time-dependent Schr\"odinger equation
\begin{align}
    \label{eq: evo psi}
    i\hbar \pdv{\ket{\Psi(t)}}{t} = H(t) \ket{\Psi(t)}\,.
\end{align}
For instance, we may be interested in tracking the evolution in time of the neutron and proton densities or the electromagnetic (e.g., dipole) moments of the nucleus.

Equation~\eqref{eq: evo psi} is non-trivial due to the time-dependence of $H(t)$.
Under the assumption of a weak external field, a perturbative solution to Eq.~\eqref{eq: evo psi} can be derived. At first order in $\mathcal{V}(t)$, as shown in (\cite{Fetter}), the time-evolved state is given by
\begin{align}
    \label{eq: psi time evolved}
    \ket{\Psi(t)} \approx \left[ 1 - \frac{i}{\hbar} \int dt^{\prime} \vartheta(t-t^{\prime}) \mathcal{V}_{I}(t^{\prime})  \right] e^{ - i H t /\hbar } \ket{ \Psi_0}.
\end{align}
The subscript $I$ indicates that an operator is evolved with respect to the unperturbed Hamiltonian (the so-called interaction picture),
\begin{align}
    \mathcal{V}_{I}(t) = e^{ - i H t /\hbar} \mathcal{V}(t) e^{iH t /\hbar}.
\end{align}
The step function $\vartheta(t-t')$ implies that the evolved state at time $t$ is influenced by the perturbation only at earlier times $t^\prime < t$, thus enforcing the essential physical property of causality.
We can then insert Eq.~\eqref{eq: psi time evolved} into Eq.~\eqref{eq: def delta Ot}. Keeping only the lowest-order terms, we then find
%\begin{align}
\begin{equation}
    \label{eq: delta A}
    \delta A(t) 
    %&
    = - \frac{i}{\hbar} \int dt^{\prime}  \vartheta(t-t^{\prime}) 
    \mel{ \Psi_0 }{ 
    \comm{ A_{I}(t) }{ \mathcal{V}_{I}(t^{\prime})} 
    }{ \Psi_0 } 
    %\\&
    \equiv \int d t^{\prime} \Pi_{AB}(t - t^\prime) v(t^{\prime}).
\end{equation}
%    \nonumber
%\end{align}
In the last line, we have introduced the so-called retarded polarization propagator\footnote{Sometimes, $\Pi_{AB}$ is denoted as the response function. However, to avoid confusion, we use this term to denote exclusively the imaginary part of $\Pi_{AB}(\omega)$, also known as the dynamical structure factor. 
}, defined as
\begin{align}
    \label{eq: polarization real time}
    \Pi_{AB}(t - t^\prime) = - \frac{i}{\hbar} \vartheta(t-t^{\prime}) \mel{ \Psi_0 }{ 
    \comm{ A_{I}(t) }{ B_{I}(t^{\prime})} 
    }{ \Psi_0 }.
\end{align}
Equation~\eqref{eq: delta A} has a clear interpretation.
At linear order, the fluctuation of any observable induced by the perturbation is determined by a convolution between the polarization and the external field. 
Importantly, $\Pi_{AB}$ depends exclusively on the properties of the unperturbed system. In fact, Eq.~\eqref{eq: polarization real time} involves the expectation value on the unperturbed ground state~$\ket{\Psi_0}$ and includes no reference to the external field.
Thus, within the linear response regime, there is a clear factorization between the probe and the internal properties of the system, which are encoded in the response function.
The polarization is most conveniently discussed in the energy representation, $\Pi_{AB}(\omega)$. By Fourier-transforming Eq.~\eqref{eq: delta A}, one finds
\begin{align}
    \delta A(\omega) = \Pi_{AB}(\omega) v(\omega),
\end{align}
where the time convolution translates into the product of the Fourier transforms of $v$, $\delta A$, and $\Pi_{AB}$.
If we then insert a complete set of eigenstates of $H$, we find that $\Pi_{AB}(\omega)$ satisfies (\cite{Fetter,giuliani_vignale_2005})
\begin{align}
    \label{eq: Polarization spectral repr}
    \Pi_{AB}(\omega) = \sum_{\nu>0} 
    \left[
    \frac{ \mel{\Psi_0}{A}{\Psi_{\nu}} \mel{\Psi_\nu}{B}{\Psi_{0}} }{ \omega - (E_{\nu}-E_{0}) + i\eta} -
    \frac{ \mel{\Psi_0}{B}{\Psi_{\nu}} \mel{\Psi_\nu}{A}{\Psi_{0}} }{ \omega + (E_{\nu}-E_{0}) + i\eta}
    \right]\,,
\end{align}
which is known as the spectral or Lehmann representation of the polarization.
There, $E_{\nu}$ are the energies of the states $\ket{\Psi_\nu}$ and $\eta$ is a small damping  parameter ($\eta \to 0^{+}$). 
The importance of Eq.~\eqref{eq: Polarization spectral repr} cannot be overestimated.
The polarization $\Pi_{AB}(\omega)$ has a specific analytical structure in the complex plane, showing (first-order) poles at the exact excitation energies of the nucleus. Also, residues are related to the transition matrix elements between the ground and excited states. 
The presence of poles means that the reaction of the system is strongest when the frequency of the external field resonantly excites a transition from the ground state to an excited state.
Therefore, Eq.~\eqref{eq: Polarization spectral repr} makes  clear that $\Pi(\omega)$ depends on the unperturbed properties of the nucleus, and conversely a measure of the linear response as a function of the frequency of the field enables us to determine the excitation energies of the system (\cite{giuliani_vignale_2005}).
Note that, as mentioned in Sec.~\ref{sec: definition response}, the completeness relation involves contributions from the full spectrum of excited states, thus including both bound states and a continuum of scattering states.

We now focus on the positive-frequency component of the polarization, defined as
\begin{align}
    \Pi_{AB}^{+}(\omega) = \sum_{\nu>0} 
    \frac{ \mel{\Psi_0}{A}{\Psi_{\nu}} \mel{\Psi_\nu}{B}{\Psi_{0}} }{ \omega - (E_{\nu}-E_{0}) + i\eta}.
\end{align}
$\Pi_{AB}(\omega)$ can be fully reconstructed from $\Pi_{AB}^{+}(\omega)$ alone.
In fact, the analytical structure of $\Pi_{AB}(\omega)$ implies that it can be reconstructed fully in terms of its imaginary part, using the so-called Kramers-Kr\"{o}nig relations, as discussed, for instance, by~\cite{Fetter} and~\cite{Fabrizio2022}.
It can be shown that
\begin{align}
    \Pi_{AB}^{+}(\omega) = - \frac{1}{\pi} \int d\omega^{\prime} \frac{
    \Im \Pi_{AB}^{+}(\omega^{\prime}) 
    }{ \omega - \omega^{\prime} },
\end{align}
which leads us to define the (linear) response function $R_{AB}(\omega)$ as
\begin{align}
    \label{eq: response function}
    R_{AB}(\omega) = -\frac{1}{\pi} \Im \Pi_{AB}^{+}(\omega)  = \sum_{\nu>0} \mel{\Psi_0}{A}{\Psi_{\nu}} \mel{\Psi_\nu}{B}{\Psi_{0}} \delta(E_{\nu} - E_{0} - \omega).
\end{align}
In the special case $A=B=O$, we recover (within a numerical factor) the definition provided in Eq.~\eqref{eq: def response function}. 
We stress again that the response function Eq.~\eqref{eq: response function} fully characterizes the reaction of a system at linear order.

 It is often useful to consider energy-weighted integrals of the response function $R(\omega)$, where the integrand is multiplied by integer powers of $\omega$. These quantities define the so-called moments of the response function. More precisely, the moment of order $k$ is obtained by weighting Eq.~\eqref{eq: def response function} with 
 the power $k$ of $\omega$,
  and integrating over $\omega$, i.e.,
\begin{equation}
\label{eq:moments_general}
    m_k=\int_{0}^\infty \, d\omega \, \omega^{k} R(\omega) =\sum_\nu|\braket{\Psi_0|O|\Psi_\nu}|^2(E_\nu-E_0)^k\,.
\end{equation}
While Eq.~\eqref{eq:moments_general} requires the knowledge of the entire spectrum of $H$, it is also possible to show that some moments can be evaluated as expectation value of operators over the nuclear ground state. This technique is often referred to as \textit{sum rules} method, and has been extensively discussed in the frame of phenomenological calculations (\cite{Bohigas78a,Orlandini91a}), with recent developments in the frame of ab initio calculations as well, see Refs.~(\cite{Porro2025,Bonaiti25Monopole}) and Sec.~\ref{sec: Monopole response}.

To summarize, we have motivated the concept of the response function on general grounds using the framework of linear response theory.
Response functions describe how a system reacts when subjected to the action of a perturbing field, such as an experimental probe. 
Under the assumption that the interaction is weak, the contributions of the perturbation and of the structure of the probed system can be factorized, and information on the internal structure is encoded fully in the first-order response $R_{AB}(\omega)$ of Eq.~\eqref{eq: response function} or $R({\omega})$ in Eq.~\eqref{eq: def response function}, when $A=B=O$. In this  Encyclopedia Chapter, we will deal only with the latter.

\subsection{Connection to experiment}
\label{sec: connection exp}
In the previous subsection, we showed that the strength function fully characterizes the linear response of the nucleus to an external probe.
Here, we connect the response functions to the evaluation of the experimentally-relevant cross sections and specify the choice of excitation operators to describe nuclear resonances. 
The latter depends on the excitation mode one aims at describing, as we now discuss.

\begin{figure}[h!]
    \centering
    \includegraphics[width=0.7\linewidth]{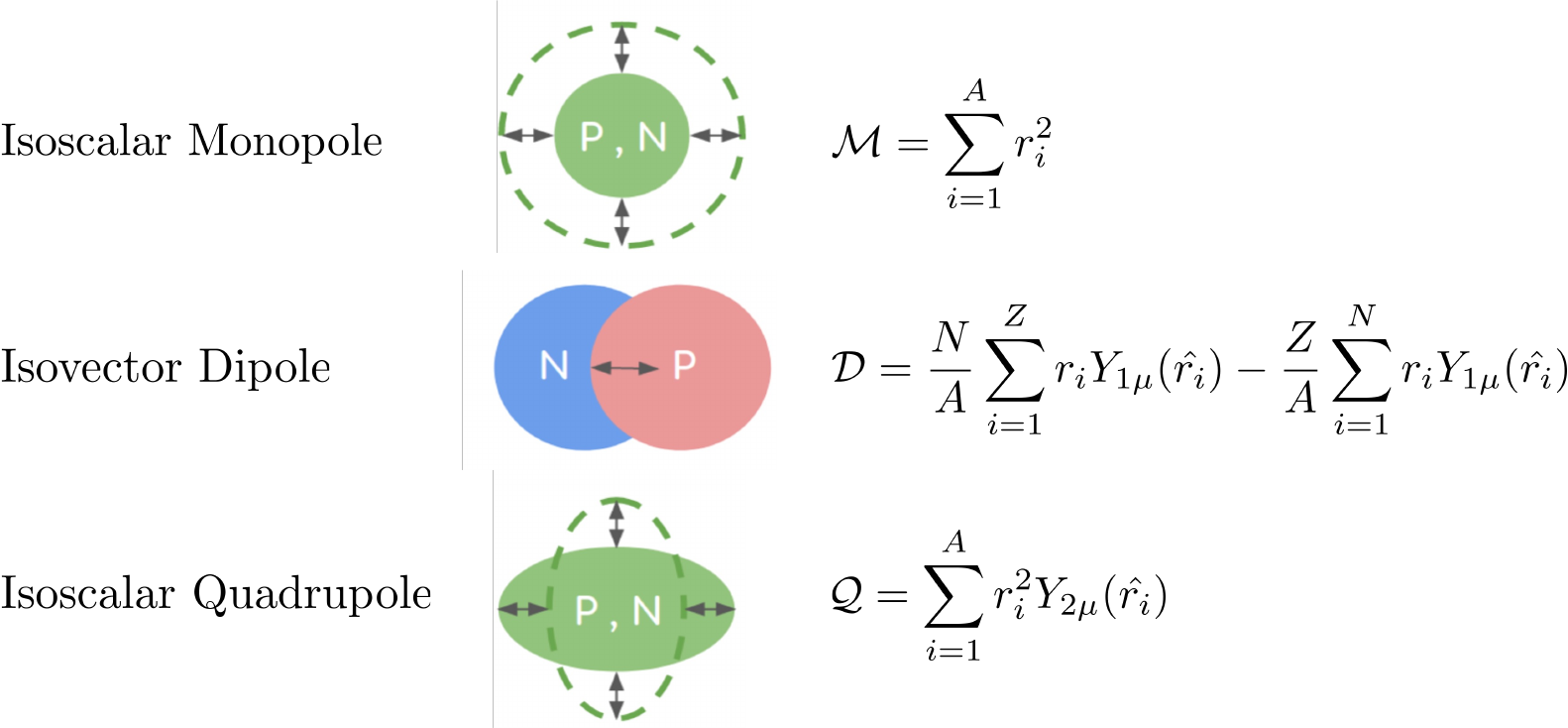}
    \caption{Schematic representation of the lowest-multipole collective modes in nuclei. The isoscalar resonances can be pictured in terms of coherent oscillations of protons and neutrons together, while isovector resonances display protons and neutrons oscillating against each other.  Corresponding expressions of the excitation operators are also presented, where $r_i$ and $\hat{r}_i$ are the absolute value and the angles of the $i$-th particle coordinate, respectively. Note that the isovector dipole operator is written in the intrinsic reference system, while the isoscalar monopole and quadrupole operators are expressed in the laboratory frame.
    }
    \label{fig:res_scheme}
\end{figure}

The main nuclear collective modes, namely the isoscalar monopole, isovector dipole, and isoscalar quadrupole, are shown pictorially in Fig.~\ref{fig:res_scheme}, where we also report the typical form of the excitation operator in terms of the nucleonic degrees of freedom.

GDRs are the most extensively studied nuclear excitations. They are typically investigated via photo‑nuclear reactions, in which a real photon—whose polarization is necessarily transverse to its direction of motion—is absorbed by the nucleus, inducing an excitation.
In the low-energy limit, after a multipole decomposition of the electromagnetic transverse current operator (\cite{Ring80a}) upon using the Siegert theorem (\cite{Siegert}), the photoabsorption cross section can be written as
\begin{equation}
\label{cs_siegert}
%\sigma_{\gamma}(\omega)=4\pi^2\alpha~\omega~R(\omega)\,,
\sigma_{\gamma}(\omega)=\frac{16\pi^3}{9}\alpha~\omega~R(\omega)\,,
\end{equation} 
where, in the long wavelength approximation, $R(\omega)$ is the response to the isovector dipole operator $\mathcal{D}$ reported in Fig.~\ref{fig:res_scheme} 
\footnote{Eq.~\eqref{cs_siegert} can be written equivalently as $\sigma_{\gamma}(\omega)=4\pi^2\alpha~\omega~R_{D_z}(\omega)$, where $R_{D_z}$ denotes the response function associated with the operator $D_z = \sum_{i=1}^{Z} (z_i - z_{cm})$, with $z_{cm}$ being the center-of-mass position. This is the convention of, e.g., (\cite{Ring80a,Bacca2013,BaccaPastore2014}). }.
Thus, experiments exploiting real photons as probes are sensitive almost exclusively to dipole excitations and allow the extraction of the nuclear response in a model-independent way (\cite{Savran2013,pietralla2019photonuclear,Zilges2022}).

A nucleus can also be probed electromagnetically in experiments where \textit{virtual} photons are exchanged.
This is the case of Coulomb excitations, where a charged projectile transfers energy to the target nucleus.
Coulomb excitation experiments at relativistic energies, in particular, are most relevant for investigating nuclear collective excitations (\cite{BertulaniBaur,Savran2013}).
Pioneering measurements of the dipole response of unstable isotopes, which can not be otherwise probed due to their short half-lives, could be performed with this technique (\cite{O22response,Aumann_2013,Savran2013}).
Relativistic proton inelastic scattering at very forward angles is also dominated by electric dipole transitions, with relatively small contributions from the nuclear interactions and from other multipolarities (\cite{VonNeumann-Cosel2019,vonneumanncosel2025electricdipolepolarizabilityconstraints}), and has demonstrated high sensitivity to the dipole strength both below and above the neutron separation energy, allowing for the extraction of the electric dipole polarizability in $^{208}$Pb (\cite{Tamii:2011pv}) and  $^{40,48}$Ca (\cite{Birkhan2017,Fearick2023}). 

Isoscalar monopole and quadrupole resonances can be probed by the inelastic scattering of $\alpha$ particles at small angles (\cite{Garg18a,Garg2023}).
The interpretation of hadronic processes is model-dependent, since they are mediated by the combined electromagnetic and nuclear interaction.
These experiments typically measure the double differential cross-section, resolved in both the momentum and the energy of the probe. For an inclusive process, where only the scattered projectile is detected but no specific final state of the target nucleus is selected, the double differential cross section reads (\cite{Bertulani2023})
%\begin{align}
\begin{equation}
    \label{eq: inelastic cross section}
    \frac{d^{2} \sigma}{d\Omega \, d\omega} 
    %&
    = \frac{k_f}{k_i} \sum_{f} \abs{ f(\theta)}^{2} \delta(E_f - E_i - \omega) 
    %\\
    %&
    = \frac{k_f}{k_i} \frac{m^2}{4\pi^2\hbar^4} \sum_{f} \abs{ \int d\mathbf{r}\chi_{f}^*(\mathbf{r})\braket{\Psi_f| \mathcal{V} |\Psi_i}\chi_i(\mathbf{r}) }^{2} \delta(E_f - E_i - \omega)\,.
    \end{equation}
    %\end{align}
Here, $k_i$ and $k_f$ denote initial and final momenta of the probe, $\chi_i$ and $\chi_f$ describe the relative projectile-target motion in the initial and final channels, respectively, the projectile-target interaction is represented by $\mathcal{V}$, the scattering angle is defined by $\cos \theta = \mathbf{k}_i \cdot \mathbf{k}_f /(k_i k_f)$, and $f(\theta)$ is the scattering amplitude for the transition to a specific state $f$.
The coordinate $\mathbf{r}$ is the relative projectile-target position, while $m$ is the reduced mass. 
As in Eq.~\eqref{eq: def response function}, the sum extends over all the internal excited states of the target nucleus, initially in $\ket{\Psi_i}$ (usually the ground state), and the Dirac delta ensures the conservation  of energy. 
Realistic calculations of Eq.~\eqref{eq: inelastic cross section} for inelastic nuclear scattering are usually performed within the distorted wave Born approximation (\cite{Satchler1964,HarakehvanDerWoude}), where $\mathcal{V}$ is described by an optical potential constrained to elastic scattering data and the $\chi_i$, $\chi_f$ wave functions are solutions to the Schr\"{o}dinger equation for the optical potential.

In the following, we show how the response functions for the multipole operators enter Eq.~\eqref{eq: inelastic cross section} under some simplifying assumptions (\cite{Pinkston61a,Colo2022Handbook})\footnote{We would like to stress that, in the case of electromagnetic (or in general electroweak) inelastic scattering, the connection between the cross sections and the nuclear response functions is essentially exact, see (\cite{WaleckaNuclearPhysics,BaccaPastore2014}. In contrast, it is only approximate in the case of hadronic scattering, since the use of the plane-wave approximation is less justified and the nuclear interaction more complex.}.
First, we approximate the short-ranged nuclear potential $\mathcal{V}$ as a contact interaction.
Also, we assume $k_i\approx k_f$, which holds for the high momenta used in $\alpha$ scattering. 
Finally, we treat the wave functions as plane waves, $\chi_i(\mathbf{r}) \sim e^{i \mathbf{k}_{i} \cdot \mathbf{r} } $, thus ignoring the distortion due to the potential.
Then, we find
%\begin{align}
\begin{equation}
    f(\theta)
    %&
    =-\frac{m}{2\pi\hbar^2}\int d\mathbf{r}e^{i\mathbf{q}\cdot\mathbf{r}}\braket{\Psi_f|\mathcal{V}|\Psi_i}
    %\nonumber\\&
    =-\frac{2m}{\hbar^2}\sum_{\lambda\mu}i^\lambda\int d\mathbf{r} \, j_{\lambda}(qr)Y_{\lambda\mu}(\hat{r})Y_{\lambda\mu}^*(\hat{q})\braket{\Psi_i|\mathcal{V}|\Psi_f}\,,
\end{equation}
%\end{align}
where $\mathbf{q} = \mathbf{k}_{f} - \mathbf{k}_{i}$ is the momentum transfer, and the plane wave has been expanded in spherical partial waves (\cite{Varshalovich88a})), where $Y_{\lambda\mu}$ denotes the spherical harmonics, $j_{\lambda}$ the spherical Bessel functions, and the magnitude and direction of $\mathbf{q}$ are $q$ and $\hat{q}$, respectively.
Similarly, the contact interaction $\mathcal{V}$ can be expanded into its multipole components,
\begin{equation}
     \mathcal{V} =V_0\sum_i\delta(\mathbf{r}-\mathbf{r}_i)=V_0\sum_i\sum_{\lambda\mu}\frac{\delta(r - r_i)}{r^2}Y_{\lambda\mu}(\hat{r}_i)Y_{\lambda\mu}^*(\hat{r}),\,
\end{equation}
where $\mathbf{r}_{i}$ are the positions of the $A$ nucleons.
In this way, one eventually obtains
\begin{equation}
    \label{eq: inelastic cross section hadronic}
    \frac{d^2\sigma}{d\Omega\,d\omega}=\left(\frac{2mV_0}{\hbar^2}\right)^2
    \sum_{f}
    \sum_{\lambda} \frac{|\braket{\Psi_f||j_\lambda Y_\lambda||\Psi_i}|^2}{4\pi}P_\lambda(\cos\theta) \, \delta(E_f - E_i - \omega) \,,
\end{equation}
where the reduced matrix element has been introduced (\cite{Varshalovich88a,Edmonds96a}) to remove the dependence on the angular momentum projection, and $P_\lambda$ is the Legendre polynomial of degree $\lambda$.
In the long-wavelength limit ($qr \ll 1$), the Bessel functions behave like $j_\lambda(qr)\approx(qr)^\lambda$, which yields the standard multipole operators definition also used in the case of electromagnetic transitions, with the difference that  when many particles are considered,  the index $i$ runs now over all the nucleons in the system, not only over protons. 
Thus, the inelastic cross section reads as a superposition of contributions of different multipolarities, each characterized by a different angular distribution,
\begin{align}
    \label{eq: multipoles inelastic scattering}
    \frac{d^2\sigma}{d\Omega\,d\omega} \propto \sum_{\lambda} R_{\lambda}(\omega) P_{\lambda}( \cos \theta),
\end{align}
where $R_{\lambda}(\omega)$ is the response associated to the operator $r^{\lambda} Y_{\lambda}$. 
The contributions of different $\lambda$'s can be disentangled thanks to their different angular distributions, allowing to extract the strength distributions from the experimental cross sections (see e.g., (\cite{Lui2001,Li:2010kfa,Garg18a,Garg2023}).
Eventually, the expressions  for the isoscalar monopole and quadrupole operators, $\mathcal{M}$ and $\mathcal{Q}$, respectively, will be as in Fig.~\ref{fig:res_scheme}.

\section{Many-body methods}
\label{sec: Many-body methods}

Computing response functions is a challenging theoretical problem, which involves determining the excitation spectrum of a nucleus up to high energies.
A variety of approaches have been devised for this problem. Methods based on the EDF framework and its extensions have long been applied to this purpose. These techniques are discussed at length in several books and reviews, for example, Refs.~(\cite{Ring80a,Colo2013,Colo2022Handbook,LiangLitvinova}).
Ab initio calculations of nuclear response functions were traditionally restricted to light nuclei (see~\cite{Efros_2007,Leidemann2013,BaccaPastore2014}).
Breakthroughs over the last decade have then made it possible to extend the reach of ab initio theory up to the medium-mass regime.

It is instructive to first set the stage by introducing the equation-of-motion (EOM) method as a general and flexible framework to tackle the excited-states problem  (Sec.~\ref{sec: Equation-of-motion method}).
The EOM allows for a compact derivation of RPA (Sec.~\ref{sec: Random Phase Approximation}), possibly the best-known theoretical approach for giant resonances, both in EDF (\cite{Colo2022Handbook}) and as a lowest-order approximation in  ab initio computations.
The core of this section is then devoted to reviewing three ab initio approaches for describing the response of medium-mass nuclei: these are coupled-cluster theory combined with the Lorentz integral transform technique (Sec.~\ref{sec: lit cc}), the projected generator coordinate method (Sec.~\ref{sec:GCM}), and self-consistent Green's functions theory (Sec.~\ref{sec: green functions}).
Other developments, such as the no-core shell model and in-medium similarity renormalization group, are also briefly discussed (Sec.~\ref{sec: other methods}).

\subsection{The equation-of-motion method}
\label{sec: Equation-of-motion method}

We introduce the EOM framework following mostly the textbooks by \cite{Ring80a,Rowe10a,Suhonen07a}.
Let us start by considering the nuclear ground state $\ket{\Psi_0}$, which is the lowest-energy solution of Eq.~\eqref{eq:Lipp}. It is possible to define (formally) a set of excitation operators $Q_\nu^\dagger$ operating on $\ket{\Psi_0}$ that create any excited state $\ket{\Psi_\nu}$ such that, for all $\nu$,
\begin{equation}
\label{eq:exc_op}
    % \begin{align}
        Q_\nu^\dagger\ket{\Psi_0}=\ket{\Psi_\nu}\,,\quad\quad
        Q_\nu\ket{\Psi_0}=0\,.
    % \end{align}
\end{equation}
For instance, one possibility for such operators would be
\begin{equation}
    Q_\nu^\dagger=\ket{\Psi_\nu}\bra{\Psi_0}\,,
\end{equation}
which is defined directly in the full \textit{infinite-dimensional} Hilbert space. However, a continuum of operators exists satisfying Eqs.~\eqref{eq:exc_op}. For example, the set of operators
\begin{equation}
Q_\nu^\dagger=\ket{\Psi_\nu}\bra{\Psi_0}+\sum_{\alpha,\beta\neq0,\nu}C_{\alpha\beta}\ket{\Psi_\alpha}\bra{\Psi_\beta}
\end{equation}
for arbitrary coefficients $C_{\alpha\beta}$ also satisfies Eqs.\eqref{eq:exc_op}. This lack of uniqueness should not discourage us from attempting to find a solution to Eqs.~\eqref{eq:exc_op}. On the contrary, it is much to our advantage, as it increases the chances of finding a suitable operator within a \textit{finite} basis of operators. From the simple definition of $Q_\nu^\dagger$ given in Eqs.~\eqref{eq:exc_op} one can easily derive the EOM for the excitation operators,
\begin{equation}
\label{eq:eom}
    [H,Q_\nu^\dagger]\ket{\Psi_0}=(E_\nu-E_0)Q_\nu^\dagger\ket{\Psi_0}
    \equiv\hbar\omega_\nu Q_\nu^\dagger\ket{\Psi_0}\,.
\end{equation}
Equation~\eqref{eq:eom} is the basic equation to solve in order to obtain the spectrum of the excited states starting from the knowledge of the ground state. 
When looking for a solution to Eq.~\eqref{eq:eom}, the first and essential step is to represent the excitation operators on a finite basis including $N$ operators $\delta Q_a^\dagger$.
The excitation operators $Q_\nu^\dagger$, then, can be expanded in this basis as
\begin{equation}
    Q_\nu^\dagger=\sum_\alpha c_\nu^\alpha \delta Q_\alpha^\dagger\,,
\end{equation}
where the linear coefficients $c_\nu^\alpha$ are still unknown. They are determined by recurring to the variational principle for the energies, namely, 
\begin{equation}
    \delta (\hbar\omega_\nu) =\delta\frac{\braket{\Psi_0|Q_\nu[H,Q_\nu^\dagger]|\Psi_0}}{\braket{\Psi_0|Q_\nu Q_\nu^\dagger|\Psi_0}}=0\,.
\end{equation}
The variation is performed with respect to the unknown coefficients $c_\nu^\alpha$, eventually giving the generalized eigenvalue problem
\begin{equation}
    \braket{\Psi_0|\delta Q_\alpha[H,Q_\nu^\dagger]|\Psi_0}=\hbar\omega_\nu \braket{\Psi_0|\delta Q_\alpha Q_\nu^\dagger |\Psi_0}\,,
\end{equation}
whose solution delivers the coefficients $c_\nu^\alpha$ for all the $Q_\nu^\dagger$ operators allowed by the spanned Hilbert subspace, as well as the corresponding eigenvalues.

Excitation operators $Q_{\nu}$ divide into two categories, Bose-like and Fermi-like excitation, obeying the Bose and Fermi canonical commutation rules, respectively. 
Thus, by exploiting Eq.~\eqref{eq:exc_op} and the canonical anti-commutation relations $\{ Q_\mu, Q_{\nu}^{\dagger} \} = \delta_{\mu\nu}$, we find 
\begin{align}
    \braket{\Psi_0|\{\delta Q_\alpha,[H,Q_\nu^\dagger]\}|\Psi_0}&=\hbar\omega_\nu \braket{\Psi_0|\{\delta Q_\alpha,Q_\nu^\dagger\}|\Psi_0}\,,
    \label{eq:eom_fermi}
\end{align}
for Fermions.
Similarly, using the commutation relations $[ Q_{\mu}, Q_{\nu}^{\dagger} ] = \delta_{\mu\nu}$, we find for Bosonic excitations
\begin{align}
        \braket{\Psi_0|[\delta Q_\alpha,[H,Q_\nu^\dagger]]|\Psi_0}&=\hbar\omega_\nu \braket{\Psi_0|[\delta Q_\alpha,Q_\nu^\dagger]|\Psi_0} \,.\label{eq:eom_bose}
\end{align}
These equations are easier to solve, because they involve evaluating ground-state expectation value of operators of a lower particle rank. 
%the object whose ground-state expectation value is to be evaluated are of a lower particle rank. 
Collective excitations are described in terms of Bose-like operators (see, for instance,~\cite{Rowe10a}), such that in the following we will refer mostly to Eq.~\eqref{eq:eom_bose}. The above equations are very flexible, and according to the choice of i) the ground-state wavefunction and ii) the spanned operator basis for the excited states, they generate a whole variety of different many-body theories. In the next section, we focus on the theory which has most widely been used for the description of nuclear excitations, namely the random phase approximation, also introduced in~(\cite{ChapterCoello}).
%\textcolor{red}{Coello Perez's contribution to this Encyclopedia}.

\subsection{The random phase approximation}
\label{sec: Random Phase Approximation}

As it was briefly mentioned in the previous section, Eqs.~\eqref{eq:eom_fermi} and~\eqref{eq:eom_bose} can generate different many-body theories depending on the particular choice of the ground state wave function $\ket{\Psi_0}$ and the operator basis spanning the excitation operators. 
The simplest choice for the wave function is given by using the Hartree-Fock (HF) ground state, where the many-body wave function is obtained by minimizing the energy within the restricted space of Slater determinants.
In this case, the many-nucleon ground state is an anti-symmetrized product of single-particle wave functions, which describes a set of independent particles moving in an average field, determined by the nucleons themselves.
The HF wave function for a nucleus of mass $A$ is obtained by filling up the $A$ lowest-lying energy levels of the single-particle potential consistently with the Pauli exclusion principle.
One refers to the single-particle states that are occupied (unoccupied) within the HF wave function as hole (particle) states. Extensive discussion about the HF mean-field can be found in textbooks (\cite{Ring80a,Blaizot1980}).
In terms of creation and annihilation operators, $c^{\dagger}_{\alpha}$ and $c_{\alpha}$, respectively, the HF ground state can be written compactly as
\begin{align}
    \label{eq: HF determinant}
    \ket{\Phi_0}  \equiv \prod_{h=1}^A c^\dagger_h \ket{0},
\end{align}
with $\ket{0}$ being the vacuum state and $c_{h}^{\dagger}$ creates a hole state in the $h$ orbit.

 \begin{figure}
    \centering\includegraphics[width=0.6\columnwidth]{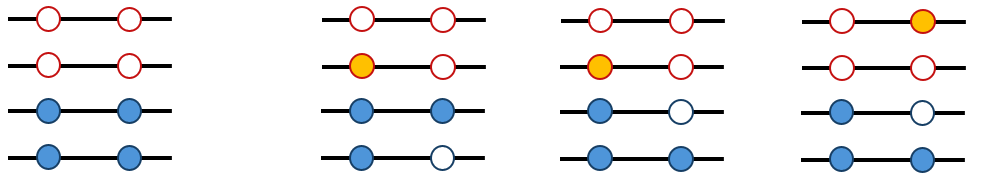}
    \caption{(Left) Pictorial representation of the Hartree-Fock ground state, obtained by filling the hole states (blue circles).
    (Right) Three examples of \NpartNhole{1} excited states obtained by promoting a nucleon from a hole state to a particle state (orange circles).
    }
    \label{fig:particle hole}
\end{figure}

Excited states of the many-body system are generated by exciting one or many individual nucleons from the hole states to particle states. 
We refer to these states as $n$-particle-$n$-hole (\NpartNhole{$n$}) states, where $n$ is the number of nucleons that are simultaneously excited. The simplest possible choice for a basis expansion of the excitation operators $Q_\nu^\dagger$ is assuming that they are built as a linear combination of \NpartNhole{1} states
\begin{equation}
\label{eq:anz_tda}
Q_\nu^\dagger\equiv\sum_{ph}c_\nu^{ph}c_p^\dagger c_h\, .
\end{equation}
%where $c_h$ annihilates a hole state in the $h$ orbit while a particle state is created by $c_p^\dagger$ in the $p$ orbit. 
A pictorial representation of \NpartNhole{1} excitations on top of the HF ground state is shown in Fig.~\ref{fig:particle hole}.

The linear coefficients $c_\nu^{ph}$ are determined by solving Eq.~\eqref{eq:eom_bose}, of which they represent the eigenvectors. 
This simple ph ansatz is known as the Tamm-Dankoff approximation (see, for instance,~\cite{Ring80a,Rowe10a}). This choice already provides a good-enough description of excited collective states. 
%It fails, however, at ameliorating the ground-state description, because the HF wavefunction is stable against \NpartNhole{1} excitations. 
An improved version of the Tamm-Dankoff ansatz, that effectively includes 1h-1p corrections to the HF ground state wave function, is given by the ansatz
\begin{equation}
\label{eq:ans_RPA}
    Q_\nu^\dagger\equiv\sum_{ph}X^{ph}_\nu c_p^\dagger c_h - \sum_{ph}Y_\nu^{ph} c_h^\dagger c_p\,,
\end{equation}
and is known as the random phase approximation. 
The variational parameters are now the $X$ and $Y$ amplitudes, which are referred to as \textit{forward} and \textit{backward} amplitudes, respectively. Once Eq.~\eqref{eq:ans_RPA} is inserted into Eq.~\eqref{eq:eom_bose}, it delivers the following non-Hermitian eigenvalue problem
\begin{align}
    \label{eq: RPA matrix}
    \begin{pmatrix}
        A & B \\
        -B^{*} & -A^{*}
    \end{pmatrix}
    \begin{pmatrix}
        X \\ Y
    \end{pmatrix}
    = \hbar\omega
    \begin{pmatrix}
        X \\ Y
    \end{pmatrix}\,,
\end{align}
where the $A$ and $B$ matrices are defined as 
\begin{subequations}
\label{eq: A B rpa matrices}
\begin{align}
    A_{ph,p'h'}&\equiv\braket{\Phi_0|[c^\dagger_h c_p,[H,c^\dagger_{p'} c_{h'}]]|\Phi_0}\,,\\
    B_{ph,p'h'}&\equiv-\braket{\Phi_0|[c^\dagger_h c_p,[H,c^\dagger_{h'} c_{p'}]]|\Phi_0}\,.
\end{align}
\end{subequations}
Equation~\eqref{eq:ans_RPA} is the most general ansatz for an excitation operator within the one-body operators space acting on the HF wavefunction. The RPA has been vastly used in many fields of physics. In ab initio nuclear theory its applications include the works by \cite{Papakonstantinou17a,Wu18a,Hu20a}.

In this section, we have shown only one of the many-possible derivations of the RPA equations, namely, how the RPA is obtained from the linearization of the EOM. Different strategies leading to a completely equivalent formulation include linear-response theory, the small-amplitude limit of the time-dependent HF, and the quadratic approximation to the generator coordinate method, which will be discussed in Sec.~\ref{sec:GCM}.

The ans\"atze given for the Tamm-Dankoff and the RPA in Eqs.~\eqref{eq:anz_tda} and~\eqref{eq:ans_RPA} are of course not unique, and neither is the choice of the reference state. Many other choices are possible starting from Eq.~\eqref{eq:eom_bose}, producing theories with increasing levels of complexity and sophistication. For instance, if the symmetry-conserving HF state is kept as a reference, one may also include \NpartNhole{2} excitations extending the original RPA ansatz~\eqref{eq:ans_RPA}. This delivers the so-called second-RPA, which has been used both within EDF calculations (\cite{Gambacurta10a,Gambacurta16a}) or starting from chiral Hamiltonians (\cite{Papakonstantinou09a,Papakonstantinou10a}). Similar results can also be obtained within the particle-vibration coupling method, which is close in spirit to the second-RPA (\cite{LiangLitvinova,Colo2022Handbook,Li22a,Li24a}). The great advantage of including \NpartNhole{2} excitations is that the dimension of the explored Hilbert space is greatly increased, producing richer spectra with more realistic fragmentation, describing the intrinsic decay width of continuum states. 
%This is necessary to reproduce, at least partially, the continuum physics, given that giant resonances lie above the particle-emission threshold.

\begin{figure}
    \centering
\includegraphics[width=0.5\linewidth]{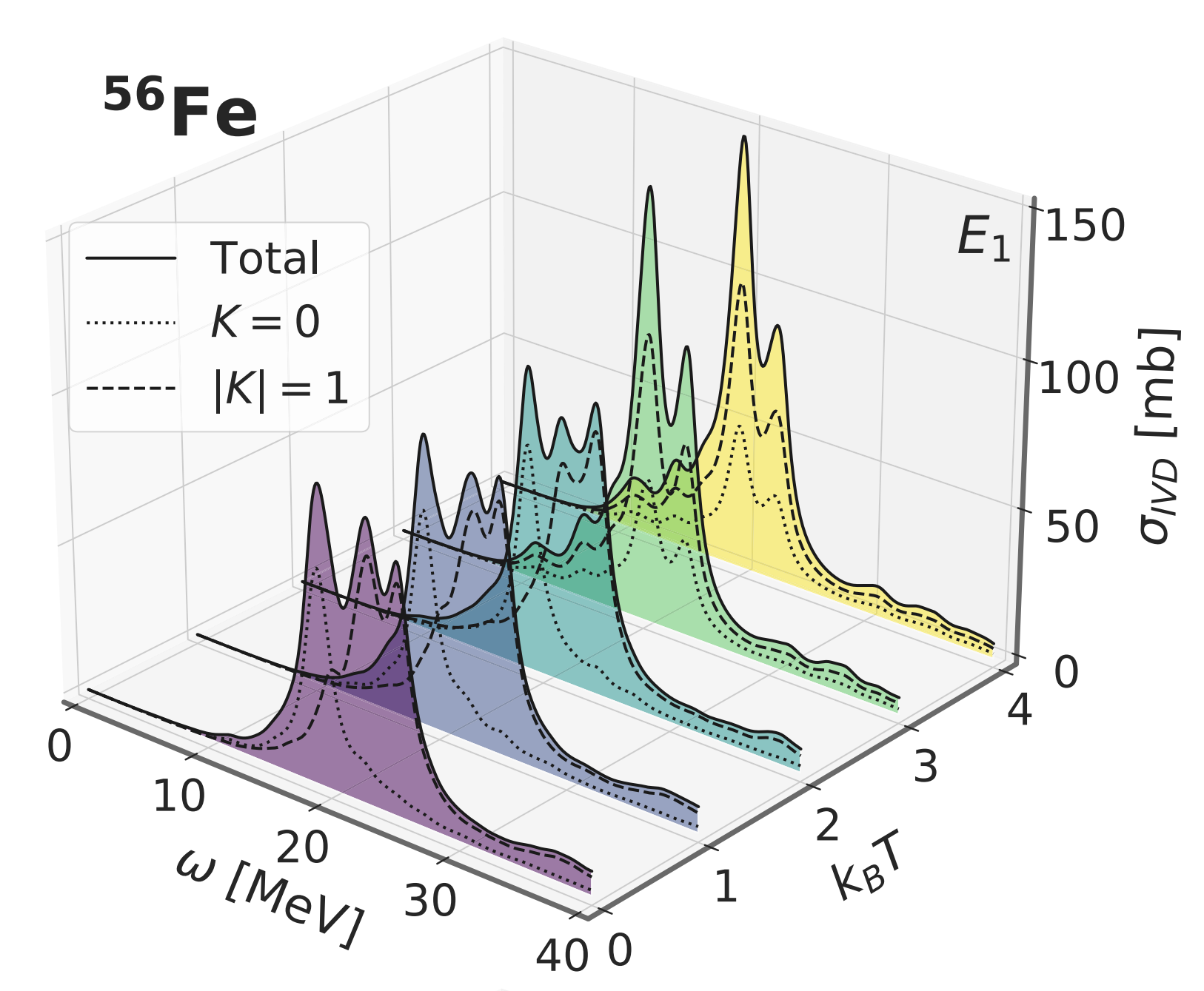}
    \caption{
    Total (isovector dipole) photoabsorption cross section  from axially-deformed QRPA calculations in $^{56}$Fe, both at zero and finite temperature, from~\cite{BeaujeaultTaudiere2022}. Different spectra are shown for different temperatures $k_\text{B}T$, displaying the evolution of the spectral shape with $T$. Curves with different $K$ represent projections of the dipole response on the $z$ axis.}
    \label{fig:fam_rpa}
\end{figure}

Symmetry-breaking reference states can also be used. This is essential in order to describe nuclei far from shell closures, where the physics of pairing and deformation plays a crucial role. Pairing is effectively taken into account by allowing the reference state to break the U(1) symmetry associated to the number of particles within the system. Particle-number-breaking reference states go under the name of Hartree-Fock-Bogoliubov (HFB) states. While they still conserve, on average, a good number of protons and neutrons, they are no longer eigenstates of the number operators $Z$ and $N$. Deformation, on the other hand, is included by allowing the reference state (either HF or HFB) to mix several total angular momenta components, such that the reference state is not an eigenstate of the angular momentum operators anymore. Further discussion about broken and restored symmetries will be addressed in Sec.~\ref{sec:GCM}.

The generalization of the RPA, whose reference state is the HFB state instead of the HF state, goes under the name of quasiparticle-RPA (QRPA). Many calculations have been systematically performed within the context of both EDFs and ab initio theories, which have allowed to understand the implications of deformation on the physics of giant resonances. Given the greater numerical cost of symmetry-breaking calculations a popular choice is represented by the Finite-Amplitude Method, first developed within the context of EDF (\cite{Nakatsukasa07a,Avogadro11a}). This method gives access to a finite resolution version of the (Q)RPA strength by generalizing the linear-response theory within the complex space (\cite{BeaujeaultTaudiere2022,Zaragoza24a}) and, while it does not directly provide the (Q)RPA wave functions, it largely reduces the numerical cost of evaluating response functions.
Example results are shown in Fig.~\ref{fig:fam_rpa}, where QRPA calculations within the Finite-Amplitude Method of the total photoabsorption cross section in the deformed nucleus $^{56}$Fe are displayed, both at zero and finite temperature (\cite{BeaujeaultTaudiere2022}).

\subsection{The Lorentz integral transform coupled-cluster approach}
\label{sec: lit cc}

Exact calculations of the open channels that enter the definition of the response functions, Eq.~\eqref{eq: def response function}, are only feasible for up to five nucleons (\cite{LAZAUSKAS2019335}) due to the difficulty in implementing scattering boundary conditions (\cite{Carbonell2013, Lazauskas2020}).
%Hence, all the ab initio techniques reviewed in this chapter require some kind of approximation of the continuum spectrum in terms of bound-state wave functions.
Integral transform approaches, in general, address these challenges by trading the computation of the response function Eq.~\eqref{eq: def response function}, with its complicated spectrum of unbound states, with that of its convolution with a properly chosen integral kernel. 
By exploiting the closure relation, one is left with the task of determining a ground-state expectation value, significantly reducing the complexity of the problem.
The integral transform can be calculated with a bound-state method of choice, and the original response function is then reconstructed by a numerical inversion.
For example, Laplace transforms have been used within imaginary-time Quantum Monte Carlo methods (\cite{Carlson:2014vla,Lynn:2019rdt}) to access electroweak response functions of light nuclei (see e.g.~\cite{Lovato:2016gkq,Lovato2020Miniboone,Gnech:2024qru}).
The Lorentz integral transform (LIT) approach (Sec.~\ref{sec: LIT}), introduced by \cite{Efros1994}, has also been exploited in a wealth of applications in few-body systems, as demonstrated in Refs.~(\cite{Efros_2007,Leidemann2013,BaccaPastore2014}) and references therein, in combination with the hyperspherical harmonics method (\cite{Bacca2002,Gazit2006}), with the no-core shell model (\cite{StetcuNcsm2009, Quaglioni:2007eg}), and recently with neural-network quantum states (\cite{Parnes2025}).
Comparisons with few-body calculations where the final states are explicitly computed have demonstrated the high accuracy of the LIT approach, see, e.g., the case of $^{3}$H in (\cite{Golak2002}).
Over the last decade, combining the LIT with  coupled-cluster theory has allowed for pioneering computations of response functions for medium-mass nuclei, as first demonstrated for the electric dipole response of $^{16}\rm{O}$ in Refs.~(\cite{Bacca2013,Bacca2014}).
The method grants a good compromise between accuracy and computational efficiency, and has then been applied to a variety of electroweak processes, see e.g.~Refs.~(\cite{Miorelli2016,Sobczyk2021,Sobczyk:2023sxh}).
Below, we will present the method, along with some pedagogical examples to ease the understanding.

\subsubsection{The Lorentz integral transform}
\label{sec: LIT}
The key idea of LIT technique (\cite{Efros1994,Efros_2007}) is to avoid computing directly the response function $R(\omega)$, but rather focus on its integral transform $L(\sigma,\Gamma)$, which is obtained by convoluting the response with a Lorentzian kernel $K_\Gamma(\sigma,\omega)$ of centroid $\sigma$ and width $\Gamma$, namely
\begin{align}
    \label{eq: lit def}
    L(\sigma,\Gamma) = \int d\omega K_\Gamma(\sigma,\omega) R(\omega) = \frac{\Gamma}{\pi} \int d\omega \, \frac{ R(\omega) }{ (\sigma-\omega)^2 + \Gamma^2 } \,.
\end{align}
By using the completeness relation in Eq.~\eqref{eq: lit def}, the LIT can be conveniently expressed as the expectation value of an operator on the nuclear ground state, namely as 
\begin{align}
    \label{eq: lit operator}
    L(\sigma,\Gamma) &= \frac{\Gamma}{\pi}
    \bra{ \Psi_{0} }
    {O}^{\dagger}
    \frac{1}{ H - E_0 - \sigma + i\Gamma}
    \frac{1}{ H - E_0 - \sigma - i\Gamma}
    O
    \ket{ \Psi_{0} } =\frac{\Gamma}{\pi} \innerproduct*{ \Tilde{\Psi}_0(z^{*}) }{ \Tilde{\Psi}_0(z) }\,,
    %\nonumber
\end{align}
where in the last step we have introduced the auxiliary state $\ket{ \Tilde{\Psi}_0(z) }$, with $z = \sigma + E_0 + i\Gamma$, defined as solution of
\begin{align}
    \label{eq: aux state}
    \left( H - z \right) \ket{ \Tilde{\Psi}_0(z) } = O \ket{ \Psi_{0} }\,.
\end{align}
Equation~\eqref{eq: aux state} has the structure of a Sch\"odinger equation with a source term in the rhs,~where the excitation operator $O$ appears.
For any finite smearing $\Gamma$, the LIT is a finite and continuous function of $\sigma$. Hence, the auxiliary state Eq.~\eqref{eq: aux state} has a finite norm and thus must satisfy bound-state boundary conditions.
Evaluating the LIT has been reconducted to a bound-state-like problem, which can be solved in principle with any many-body method, provided that an explicit representation of the wave function is given.

Once the LIT has been determined, a further step consists in recovering $R(\omega)$.
From the mathematical point of view, an ill-posed inversion problem must be solved, since there is no analytical kernel for the inverse transform.
Indeed,  the Lorentz kernel smears the structure of the response function and is always computed within a certain numerical error. Therefore,  different response functions may exist that produce similar LITs within such numerical error,
so that some care and expert knowledge are needed to get robust predictions of $R(\omega)$.
The problem of inverting the LIT is reviewed in detail in Refs.~(\cite{Efros_2007,Barnea:2010kcs}).  The adopted inversion method leads to a smooth response function, where the continuum is recovered.

\subsubsection{Coupled-cluster theory}
\label{sec: cc theory}
Coupled-cluster (CC) theory is a powerful ab initio method, which has featured a variety of applications in nuclear physics, including ground-state properties (energies, densities, and electroweak form factors), low-lying excited states, and, in combination with the LIT technique, response functions.
The mild computational scaling of CC theory has made it possible, in particular, to pioneer ab initio calculations of nuclear response functions in medium-mass isotopes in \cite{Bacca2013,Bacca2014}.
More recently, the scope of CC has been extended in the directions of heavy nuclei in the Pb region (\cite{PbAbInitio,Bonaiti:2025bsb}), semi-magic isotopic chains (\cite{Tichai2024}), and deformed nuclei (\cite{Hagen2022,Sun2025}).
Applications to the electric dipole response of open-shell systems in the vicinity of a shell closure have also been put forward in Refs.~\cite{Bonaiti2024,Marino2025}.
For the sake of simplicity, we will focus here on closed-shell spherical nuclei and introduce the essential features of CC theory, which is discussed in more detail in \cite{ShavittBartlett,Hagen2014Review} and in ~(\cite{ChapterFossezHergert}). %\textcolor{red}{Fossez and Hergert's contribution to the Encyclopedia}.
 
In its basic formulation, CC is a theory in which an accurate parametrization of the correlated ground state of a given nucleus is determined by the exponential ansatz
\begin{align}
    \label{eq: cc gs ansatz}
    \ket{\Psi_0} = e^{ T } \ket{ \Phi_0 },
\end{align}
where $\ket{ \Phi_0 }$ is a reference state, on top of which correlations are built by the action of $e^{ T }$, where the cluster operator $T$ is expanded as a combination of \NpartNhole{$n$} excitation operators $T_n$.
In the case of closed-shell nuclei, $\ket{ \Phi_0 }$ is typically taken to be a spherical HF solution.
While in principle $T$ should include all contributions up to \NpartNhole{$A$}, in all practical applications the operator is truncated up to a much smaller number of particle-hole excitations.
As we discuss below, the exponential ansatz is very efficient in encoding dynamical correlations in the wave function, yielding the bulk of the nuclear binding energy even for low orders $n$ (\cite{papenbrock2024}).
The standard truncation is called CC at the singles and doubles level (CCSD) and approximates the cluster operator as $T \approx T_1 + T_2$, where
\begin{subequations}
\begin{align}
     T_1 = \sum_{ai} t^{a}_{i} c_a^{\dagger} c_i\, \quad{\rm and}\quad
     T_2 = \frac{1}{4} \sum_{abij} t^{ab}_{ij} c_a^{\dagger} c_b^{\dagger} c_j c_i\, ,
\end{align}
\end{subequations}
and $i,j$ and $a,b$ denote hole and particle states, respectively.

The CC exponential ansatz induces a similarity (non-unitary) transformation on the Hamiltonian, which reads
\begin{align}
    \label{eq: similarity transformed H}
    \Bar{H} = e^{- T} {H}_{N}  e^{T} = ( {H}_N e^{T} )_{C}
\end{align}
where we have defined the normal-ordered operator ${H}_N$ by subtracting its expectation value on the reference $E_{0}^{\rm ref} = \mel{\Phi_0}{H}{\Phi_0}$, namely, ${H}_N = {H} - E_{0}^{\rm ref}$.
The subscript $C$ indicates that only connected contributions must be considered, as follows from applying the Hausdorff expansion to $e^{T}$, see e.g.~(\cite{Bartlett2007,Hagen2014Review}).
This is a crucial property of CC theory, which guarantees the size-extensivity of the method (namely, energies and wave functions have the correct scaling with the number of particles) and limits considerably the number of diagrams to be considered.
Inserting Eq.~\eqref{eq: cc gs ansatz} into the~\Sch equation $H \ket{\Psi_0} = E_{0} \ket{\Psi_0} $, the latter can be recast into the form
\begin{equation}
    \label{eq: CC basic eq}
    \Bar{H} \ket{ \Phi_0 } = \Delta E_0 \ket{ \Phi_0 },
\end{equation}
where $\Delta E_0 = E_{0} - E_{0}^{\rm ref}$ is the CC correlation energy.
The $T$  amplitudes are determined as the solution to a set of non-linear equations obtained by projecting Eq.~\eqref{eq: CC basic eq} onto the excited Slater determinants $\ket{ \Phi_{i}^{a} } = c_{a}^{\dagger} c_i \ket{\Phi_0} $, $\ket{ \Phi_{ij}^{ab} } = c_{a}^{\dagger} c_{b}^{\dagger} c_{j} c_{i} \ket{\Phi_0} $, and solved iteratively.

The ansatz Eq.~\eqref{eq: cc gs ansatz} allows to determine the ground state~energy.
However, an additional step is needed to access other ground state~observables. 
Since $\Bar{H}$ is non-Hermitian, the left ground state~$\bra{\Psi_0}$ is not the adjoint of $\ket{\Psi_0}$.
In the CCSD scheme, $\bra{\Psi_0}$ is parametrized in terms of a set of \NholeNpart{1} and \NholeNpart{2} de-excitation amplitudes, denoted as $\Lambda$, 
\begin{align}
    \label{eq: Left ground state}
    \bra{\Psi_0} = \bra{\Phi_0} (1 + \Lambda) e^{-T}.
\end{align}
The $\Lambda$ amplitudes (which commute with the $T$'s operators) satisfy a set of linear equations, which are solved after having determined the $T$ amplitudes.

More accurate solutions with respect to the CCSD approximation can be obtained by including triples (\NpartNhole{3}) contributions in $T$, e.g., in the approximate CCSDT-1 truncation scheme (\cite{Hagen2014Review,Miorelli2018}).
Typically, CCSD allows to recover roughly 90\% of the correlation energy, with CCSDT-1 yielding an additional 10\% correction, while in general the effect of triples on ground-state observables is modest.

\subsubsection{Combining the Lorentz integral transform and coupled-cluster theory}
\label{sec: litcc subsection}

%The starting point  is the  definition Eq.~\eqref{eq: def response function} of the response function.
%By using the completeness relation, the response function for a closed-(sub)shell nucleus ($J=0$)  can be written as the ground state expectation value 
%\begin{align}
%    R(\omega) = \mel{\Psi_{0}}{ O^{\dagger} \delta(H - E_0 - \omega) O }{ \Psi_{0}}.
%\end{align}
%Inserting Eqs.~\eqref{eq: cc gs ansatz} and~\eqref{eq: Left ground state} for the CC right and left ground state, one gets
%\begin{align}
 %   R(\omega) = \mel{ \Phi_0 }{ 
%    (1+\Lambda) \, \Bar{O}^{\dagger} \delta\left( 
%    \Bar{H} - \omega
%    \right)\Bar{O}
%    }{ \Phi_0 },
%\end{align}
%where the similarity-transformed excitation operator $\Bar{O}$ has been defined in analogy with Eq.~\eqref{eq: CC basic eq}.

The LIT approach can be formulated in the CC language by writing Eq.~\eqref{eq: lit operator} as
\begin{align}
    L(\sigma,\Gamma) = \frac{\Gamma}{\pi} \innerproduct*{ \Psi_L(z^{*}) }{ \Psi_R(z) },
\end{align} 
where the auxiliary states $\Psi_L(z^{*})$, $\Psi_R(z)$ are defined as the solutions to the Schr\"{o}dinger-like equations 
\begin{subequations}
    \label{eq: CC auxiliary}
    \begin{align}
        & \left( \Bar{H} - z \right) \ket{ \Psi_R(z) } = \Bar{O} \ket{\Phi_0}, 
    \end{align}
    \begin{align}
        & \bra{ \Psi_L(z^{*}) } \left( \Bar{H} - z^{*} \right) = \bra{\Phi_0} (1 + \Lambda) \Bar{O}^{\dagger} \,.
    \end{align}
\end{subequations}
Here, one can recognize that the CC language is used because of the presence of the  similarity-transformed Hamiltonian and the similarity-transformed excitation operator $\Bar{O}$, which is defined in analogy to Eq.~\eqref{eq: similarity transformed H}.

The workflow of LIT-CC requires solving for the $T$ and $\Lambda$ amplitudes first, which are needed to construct the similarity-transformed operators.
Then, in Eq.~\eqref{eq: CC auxiliary}, the auxiliary states are expanded, in a configuration-interaction-like way, as vectors in the space of \NpartNhole{$n$} configurations.
This is called the equation-of-motion CC (EOM-CC) ansatz and is discussed in depth in Refs.~\cite{Krylov2008,Hagen2010,Hagen2014Review}.
The choice of terminology stems from the basic assumption of the EOM-CC consists in parametrizing the excited states as the outcome of linear operators generating \NpartNhole{$n$} excitations on top of the correlated CC ground state $\ket{\Psi_0}$, as in Eq.~\eqref{eq:exc_op} and related discussion.
For our purposes, we write $\ket{ \Psi_R(z) } = \mathcal{R}(z) \ket{\Phi_0}$, where typically $\mathcal{R}(z)$ includes up to \NpartNhole{2} terms.
We refer to a computation at the \NpartNhole{2} level in both the ground state and the EOM ansatz as CCSD, which is a very good starting point. 
Triples may correct somewhat the strength distribution, typically by shifting the response function to higher energies (\cite{Miorelli2018,Marino2025,Marino2025Ischia}).
After the computation of the LIT, one inverts  the transform to retrieve the
response function $R(\omega)$.

A pedagogical example is presented below for the dipole strength function of the $^{16}$O nucleus computed with the NNLO$_{\text{sat}}$(450) interaction from~\cite{NNLOsat} using the CCSD approximation. 
\begin{figure}[h!]
    \centering
    \includegraphics[width=\columnwidth]{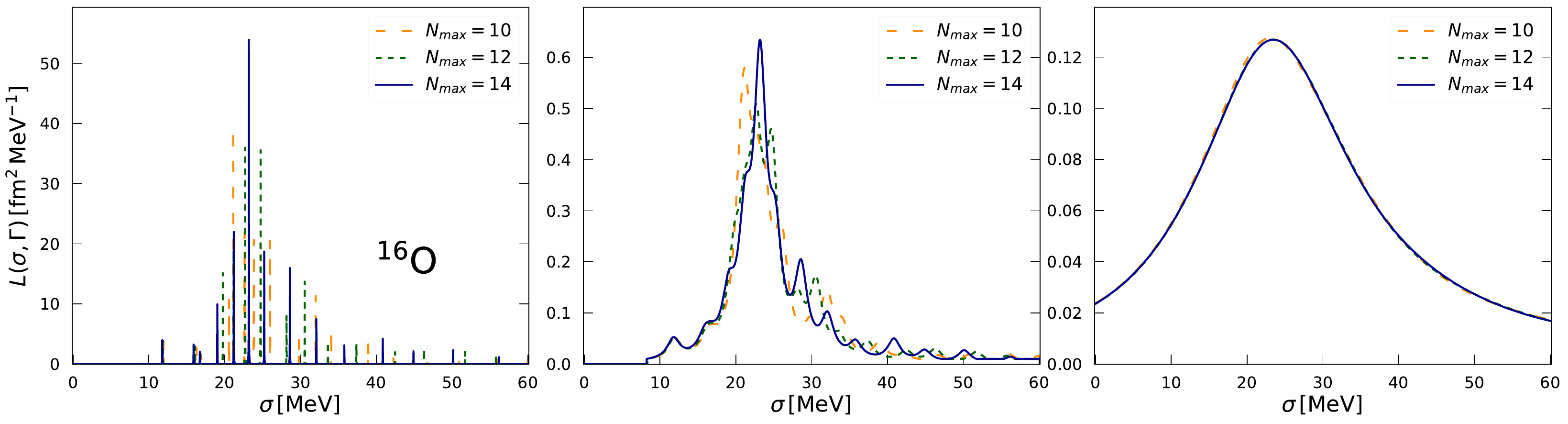}
    \caption{LIT of the dipole response for the $^{16}$O nucleus plotted for three different values of the width parameter $\Gamma=0.01, 1, 10$ MeV (from left to right). Three different model space sizes are shown in each panel (see text for details).
    All calculations are performed at a harmonic oscillator frequency $\hbar\omega = 14 \,\rm{MeV}$ employing the NNLO$_{\text{sat}}$(450)  interaction.
    }
    \label{fig:LIT_discretized}
\end{figure}
In Figure \ref{fig:LIT_discretized}, we plot the LIT of the dipole response for three different values of the width parameter $\Gamma=0.01, 1$, and $10$ MeV and different model-space dimensions $N_{max} = 10, 12, 14$, where the number of harmonic oscillator shells included is given by $N_{max}+1$.
For the  $\Gamma=0.01$ MeV case, we clearly see several discrete peaks.
As in the limit $\Gamma\rightarrow 0$ the Lorentzian kernel becomes a delta function, effectively, the leftmost panel of Fig.~\ref{fig:LIT_discretized} is nothing else than a ``discretized response", which represents a bound-state approximation to $R(\omega)$.
%, since the integral transformation only changes the variable $\omega$ to $\sigma$ in Eq.~(\eqref{eq: lit def}). 
The distribution of the peaks depends strongly on the model-space dimension, controlled by the parameter $N_{max}$.
Despite $N_{max}=14$ being the largest model space we can access, the computed discretized response is not fully converged yet. 

The second panel of Fig.~\ref{fig:LIT_discretized} shows the LIT for $\Gamma=1$ MeV. Here, one can definitely see that  the integral transform has a smearing effect. The several peaks visible in the discretized response  between 20 and 25 MeV  are broadened by the finite width $\Gamma$.
However, also in this case the calculation 
is  not yet converged in $N_{max}$. 
This curve can be interpreted as a ``discretized response folded with a Lorentzian  of $\Gamma=1$ MeV". 
A Lorentzian folding with a width comparable to the experimental resolution is often applied to response function calculations performed with RPA or other many-body techniques (see also Sec.~\ref{sec: results}).
This step is required to obtain a continuous response for comparison with experimental data. However, one has to be aware that the shape of the resulting curve may depend significantly on the value of $\Gamma$. 
For example, choosing $\Gamma=2$ MeV would differ from the curve shown for $\Gamma=1$ MeV.
The conceptual difference with the LIT approach is that, in the latter, we interpret the smeared function as an integral transform and not a response function itself.  
Still, if we attempted to invert the curve for $N_{max}=14$ in the middle of Fig.~\ref{fig:LIT_discretized}, we would not obtain any stable solution, because the numerical error of the calculation is still too large.

Finally, in the third panel of Fig.~\ref{fig:LIT_discretized}, the LIT for $\Gamma=10$ MeV is shown. For such a large width, the smearing effect is so strong that the underlying structure of the response function is washed out, and only a large bell-shaped peak in the region of the giant resonance is visible.
However, convergence with respect to the model-space size is clearly achieved, and therefore the resulting LITs can be inverted within a certain numerical accuracy. Results after the inversion are shown later in Sec.~\ref{sec: results}, where we compare to other theoretical approaches and to experimental data.

In essence, the philosophy of the LIT approach is to first compute the LIT with the highest possible precision and then invert it.
A finite $\Gamma$ (i.e., it cannot be infinitesimally small) must be taken to ensure a stable reconstruction of $R(\omega)$.
Convergence in the inversion procedure can be assessed by checking that the response functions obtained for different $\Gamma$'s within a sensible range (say, between 5 and 20 MeV) and different $N_{max}$'s lie reasonably close to each other.
Thus, while the inversion process adds complexity to the method, it also paves the way for assessing the theoretical uncertainties on the reconstructed response function and the associated cross sections.
For instance, the spread for different $\Gamma$'s can be interpreted as an error bar of the inversion.

\begin{figure}[h!]
    \centering
    \includegraphics[width=0.46\linewidth]{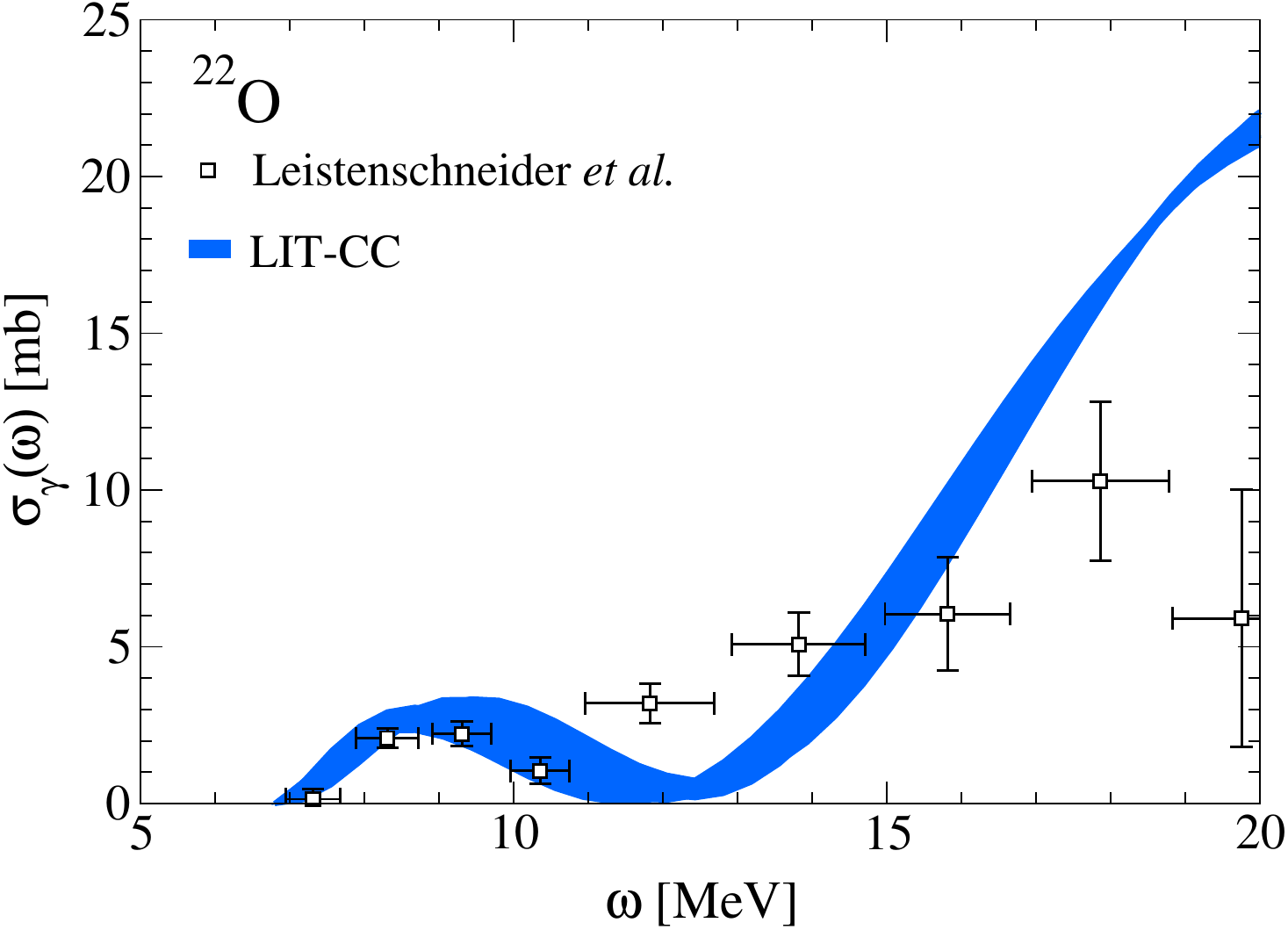}
    ~\includegraphics[width=0.45\linewidth]{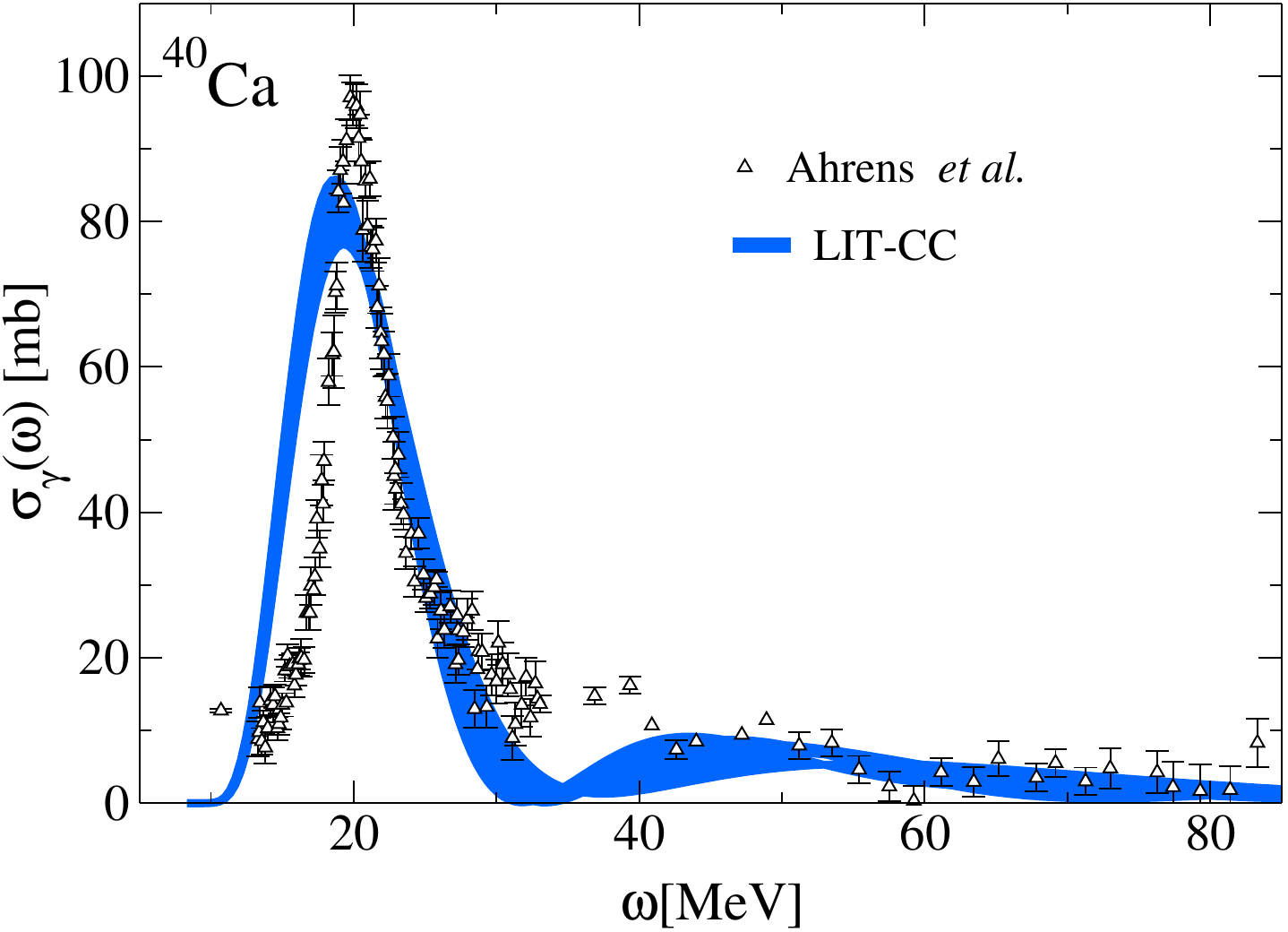}
        \caption{LIT-CC results in the CCSD approximation for the $^{22}$O (left) and $^{40}$Ca (right) photoabsorption cross section, in comparison to experimental data from \cite{O22response, AhrensO16}. 
        Error bands represent an estimate of the theoretical uncertainties associated to the inversion of the LIT.
        The N3LO chiral interaction (with two-nucleon forces only) from (\cite{EntemMachleidt}) is employed.
        Figure adapted from \cite{Bacca2014, Simonis:2019spj}. 
    }
    \label{fig:LIT-CC-results}
\end{figure}

An example of results  for the dipole response functions obtained though the  LIT-CC method after an inversion is shown in Fig.~\ref{fig:LIT-CC-results}. These calculations are based on the CCSD approximation starting from a two-body Hamiltonian at next-to-next-to-leading order (N3LO) in the chiral expansion (\cite{EntemMachleidt}). The width of the shown curves corresponds to the estimated uncertainty of the inversion procedure, with more  details found in Refs. \cite{Bacca2014,Simonis:2019spj}. For both $^{22}$O (left panel) and $^{40}$Ca (right panel), good agreement with experimental data from Coulomb excitation and photoabsorption, respectively, is observed. In particular, within this theory one observes the emergence of a pigmy resonance for the neutron-rich nucleus of $^{22}$O, and of  a giant resonance for the stable $^{40}$Ca nucleus.

Finally, one more comment is in order regarding the vanishing $\Gamma$ case.  Since in this case the Lorentz distribution reduces  to a Dirac's delta function, it is also easy to show that the moments $m_n$ of the response function can be evaluated for any order $n$ by integrating the LIT for a small value of the width, without the need of any inversion (\cite{Miorelli2016}),
\begin{align}
    \label{eq: moments def}
    m_n = \int d\omega \, \omega^{n} R(\omega) 
    = \lim_{\Gamma \to 0} \int d\sigma \, \sigma^{n} L(\sigma,\Gamma).
\end{align}
As a consequence, the LIT-CC framework is particularly efficient for evaluating electromagnetic sum rules.
In Sec.~\ref{sec: results}, we will present results for sum rules obtained with this method in comparison to other approaches. 
%sum rules can be readily evaluated within the LIT framework. 

\subsection{The generator coordinate method}
\label{sec:GCM}

Among the many-body techniques addressing the giant resonances, interest was shown in multiple exploratory works for the generator coordinate method (GCM) (\cite{Caurier73a,Abgrall75a,Flocard75a,Stoitsov94a}). However, no extensive use of the GCM to describe response functions followed these seminal efforts. 
%in using the GCM to describe GRs.
New efforts have been  taken recently in a series of work (\cite{Porro24a,Porro24b,Porro24c,Porro24d}) that have exploited the power of Projected GCM in order to describe from an ab initio standpoint the physics of the giant monopole and, to some extent, quadrupole resonances.

The projected generator coordinate method (PGCM) is a popular and versatile many-body method based on the mixing of Bogoliubov vacua typically generated by solving constrained HFB mean-field equations (\cite{Ring80a,Schunck19a,Bally24a}).
It belongs to the class of multi-reference approaches, which are able to access, in principle, all open-shell systems by efficiently capturing the static (or collective) correlations associated to deformation and pairing.
PGCM has been used prevalently in the context of EDF theory (\cite{Schunck19a}).
However, it has been extended recently to the ab initio framework (\cite{Frosini21b}), allowing for an accurate description of ground-state and lower-lying states of light- and medium-mass nuclei starting from chiral interactions (\cite{Frosini21b,Giacalone24a,Bally24a,Bally24b}).

\subsubsection{The projected generator coordinate method ansatz}
Within the GCM, the wavefunction ansatz is a general continuous superposition of so-called generating functions $\ket{\Phi(q)}$ reading (\cite{Hill52a,Griffin57a})
\begin{equation}
\label{eq:GCM_ans}
    \ket{\Psi_\nu}=\int dq f_\nu(q)\ket{\Phi(q)}\,,
\end{equation}
where $q$ is a set of collective variables referred to as \textit{generator coordinates}. 
An ensemble of non-orthogonal basis states $\{\ket{\Phi(q)}, q\in [q_0,q_1]\}$ is usually chosen to be a set of 
constrained Bogoliubov vacua, which satisfy the condition
\begin{equation}
    \braket{\Phi(q)|Q|\Phi(q)}=q\,,
\end{equation}
where $Q$ is a one-body operator.
The eigenstates of the nuclear Hamiltonian $H$ are indicated by $\ket{\Psi_\nu}$ and are represented as linear combinations of the $\ket{\Phi(q)}$'s, where $f_\nu(q)$ is a set of weight functions to be determined.
By means of the variational principle, i.e., minimizing the expectation value of the Hamiltonian $H$ within the space spanned by the basis states, 
the so-called Hill-Wheeler-Griffin (HWG) equation
\begin{equation}
\label{eq:HWG}
    \int dq'\Big[\mathcal{H}(q,q')-E_\nu\mathcal{N}(q,q')\Big]f_\nu(q')=0
\end{equation}
is obtained, where the energy and norm kernels
\begin{subequations}
\label{eqs:kernels}
    \begin{align}
        \mathcal{H}(q,q')&\equiv\braket{\Phi(q)|H|\Phi(q')}\,,\\
        \mathcal{N}(q,q')&\equiv\braket{\Phi(q)|\Phi(q')}\,,
    \end{align}
\end{subequations}
have been introduced. Equation~\eqref{eq:HWG} is a generalised eigenvalue problem for a set of non-orthogonal basis states, hence the presence of a norm matrix. The solution of Eq.~\eqref{eq:HWG} gives a spectrum of states $\ket{\Psi_\nu}$ spanning the chosen Hilbert subspace. The above equations are given for a single constraining operator $Q$, but several operators can (and usually are) constrained at the same time, so to explore a multidimensional manifold which allows to address coupling effects between different collective coordinates. The choice of the collective coordinates rests very much on prior knowledge of the physics of interest (see below).

Constrained HFB solutions typically break symmetries of the initial Hamiltonian, such that the restoration of such symmetries is mandatory to discard spurious symmetry-breaking effects. In particular, while still carrying \textit{on average} the good number of protons and neutrons $Z$ and $N$, HFB states are not eigenstates of particle numbers operators, i.e., the associated variance is nonvanishing,
\begin{subequations}
    \begin{align}
        \braket{\Phi(q)|Z^2|\Phi(q)}\neq0\,, \quad
        \braket{\Phi(q)|N^2|\Phi(q)}\neq0\,.
    \end{align}
\end{subequations}
Similarly, in order to include the effects of so-called collective correlations, 
%(of the many-particle-many-hole type)
the vacua are allowed to break spatial symmetries, such as invariance under rotation and reflection, producing wavefunctions which are \textit{not} eigenstates of the total angular momentum $J^2$, its \textit{z} projection $M$ or parity $\Pi$. The complete set of symmetry quantum numbers is specified by the index $\sigma$, defined as
\begin{equation}
    \sigma\equiv\{N,Z,J,M,\Pi\}\,.
\end{equation}
Naturally, the number of symmetries which are broken by the basis states depend on the specific GCM implementation. In order to eventually restore the symmetries of the Hamiltonian and to have solutions which carry \textit{good quantum numbers} (i.e., that are eigenstates of the proton and neutron number, angular momentum and parity operators) projection operators $P^\sigma$ associated with the group $\mathcal{G}$ (which can be both continuous or discrete) are used, which are generically written as
\begin{equation}
    P^\sigma=\int d\varphi~g^\sigma(\varphi)R(\varphi)\,.
\end{equation}
The function $g^\sigma(\varphi)$ represents irreducible representations of $\mathcal{G}$ while $R(\varphi)$ is a unitary symmetry transformation operator changing the orientation of the state by the angle $\varphi$. Thus, the symmetry-conserving version of the GCM ansatz~\eqref{eq:GCM_ans}, i.e., the  PGCM ansatz, reads
\begin{equation}
    \ket{\Psi_\nu^\sigma}\equiv\int dq f_\nu^\sigma(q)P^\sigma\ket{\Phi(q)}\,.
\end{equation}
The additional $\sigma$ dependence within the linear coefficients $f_\nu^\sigma(q)$ is resolved by noticing that the action of the projection operator $P^\sigma$ produces several separated HWG equations~\eqref{eq:HWG}, one for each symmetry quantum number set. The equations are eventually solved separately, such that different spectra for each irreducible representation (e.g., the angular momentum and parity of the target states) are provided.

Notice that PGCM does not include dynamical (beyond-HFB) correlations, at variance with the coupled-cluster or SCGF expansions. Thus, only a fraction of the total energies is typically captured (\cite{Frosini21b}). However, it is effective in describing both low-lying spectroscopy (e.g., rotational bands) and high-energy spectra (\cite{Frosini21b,Porro24a}), thanks to the substantial cancellation of dynamical effects on these observables and its ability to capture the collective properties of a nucleus.

\subsubsection{Projected generator coordinate method calculations of the giant monopole resonance}
\begin{figure}
    \centering
    \includegraphics[width=0.594\linewidth]{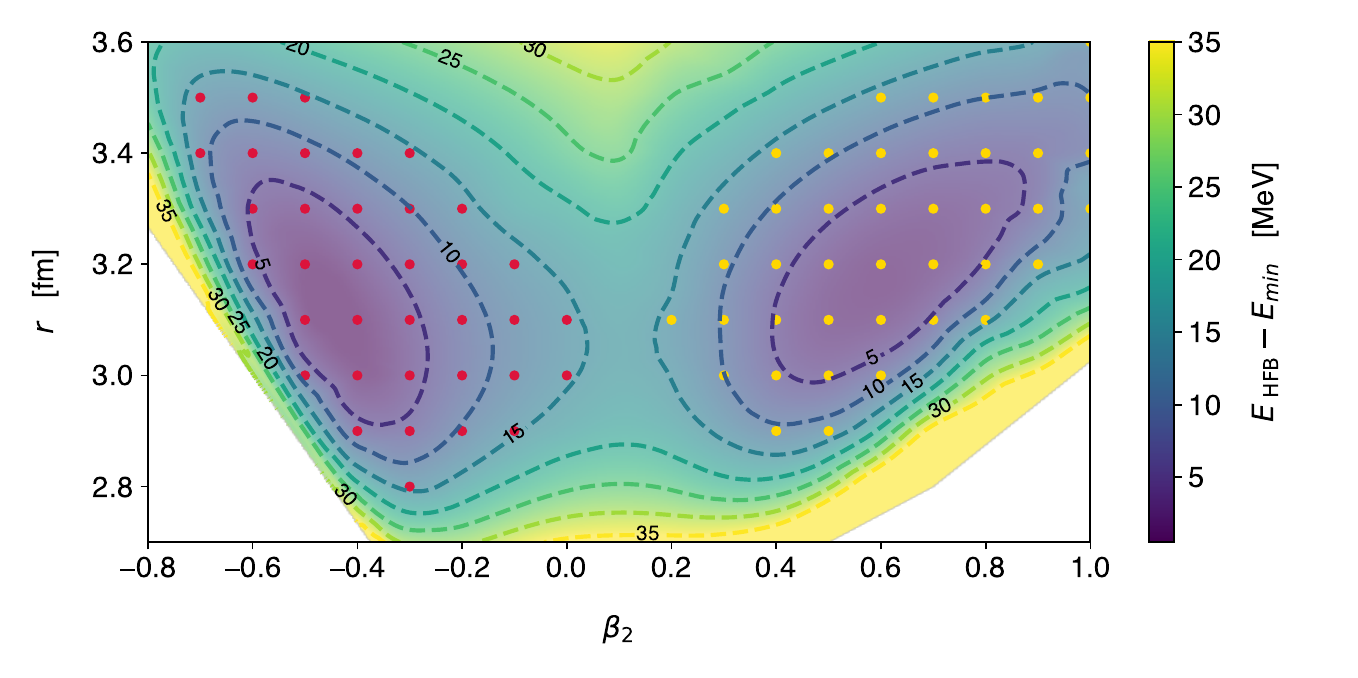}
    \includegraphics[width=0.40\linewidth]{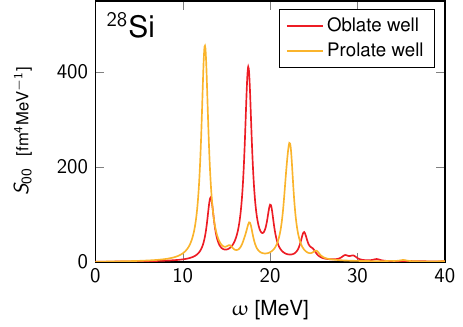}
    \caption{Left: Total energy surface at the mean-field level in $^{28}$Si. Right: isocalar monopole response in $^{28}$Si from PGCM calculations. The red curve is the result of the PGCM computation employing the oblate ansatz (red dots in the left panel), while the yellow curve shows the results for the prolate ansatz (yellow dots in the left panel). Both figures have been adapted from (\cite{Porro24b}). Calculations were performed employing the chiral potential from (\cite{Huther19a}) at N3LO in both the two- and three-nucleon sector.}
    \label{fig:HFB_PES}
\end{figure}

In order to better understand the mechanism behind the (P)GCM formalism, results for the isoscalar giant monopole resonance (GMR) in $^{28}$Si from \cite{Porro24b} are presented. Fig.~\ref{fig:HFB_PES} (left) shows the total energy surface (the energy obtained at the HFB level for each choice of the coordinates $q$) in $^{28}$Si as a function of two different generator coordinates, namely the mean square radius $r=\sqrt{\braket{r^2}}$ and the dimensionless quantity $\beta_2$, which is proportional to the axial quadrupole deformation $\braket{Q_{20}}$. Positive values of $\beta_2$ indicate a prolate shape (elongated), typically represented by a rugby ball, while negative $\beta_2$ values describe oblate shapes, i.e., compressed in the direction of a rotation axis (pancake-like). As it was mentioned in the previous discussion, the choice of the coordinates depends on the specific physics case one wants to study. In this case, since the isoscalar GMR is a shape-conserving vibration of the whole nucleus, well described by the expression \textit{breathing mode} (\cite{Garg18a}), it is natural to assume that this kind of physics can be captured by combining HFB solutions corresponding to different values of the radius.
Additionally, it is known empirically that the most important coupling effect arises from the interplay of monopole and quadrupole degrees of freedom, such that the simultaneous exploration of the $\braket{Q_{20}}$ dimension is mandatory in the physical description of deformed nuclei. Thus, the unspecified generator coordinate $q$, which was used in Eq.~\eqref{eq:GCM_ans} to label the generating functions $\ket{\Phi(q)}$, is replaced by two well-defined physical quantities, such that Fig.~\ref{fig:HFB_PES} (left) is a map of the HFB energy over the set of basis states $\ket{\Psi(r,\beta_2)}$. The choice of the generator coordinates represents the first step in a (P)GCM calculation, and it is by far the aspect of this method that is most strongly subject to prior knowledge about the physical phenomenon to be described.

The following step is represented by the selection of the discrete set of basis states to be included in the ansatz~\eqref{eq:GCM_ans}. Equation~\eqref{eq:GCM_ans} is written in its most general representation as a continuous superposition, but numerical calculations replace, by necessity, the continuous integral with a discrete sum over a set of HFB states. We may refer to this step as the choice of a discrete mesh for a given generator coordinate. The choice of the mesh and the choice of the generator coordinates are not completely decoupled issues, and stringent numerical benchmarks are needed to assure the stability of the numerical results against the discretisation of continuous variables. The interested reader is referred to the works by \cite{Martinez-Larraz22a,Porro24a,Bofos25a}. The main criterion is based on the energy difference from the minimal-energy point in the multi-dimensional manifold, since HFB states very far in energy from one another have very little overlap, and are thus not contributing when eventually solving the HWG equation. In this case (Fig.~\ref{fig:HFB_PES}) (left) two different sets of discrete HFB states are considered, entering two different ans\"atze, one only including oblate configurations (red dots) and another only in the prolate region (yellow dots).
Once the basis states have been selected, the corresponding energy and norm kernels from Eqs.~\eqref{eqs:kernels} are evaluated. This part represents the most numerically intense effort of the procedure, especially when many symmetries are broken at the HFB level. In such case symmetries need to be restored within the PGCM, and the cumulated cost of several projections may result in a significant numerical workload, the most demanding part being the restoration of rotational symmetry. 
In this example, where axial deformations were allowed, projections over $N$ (7 mesh points), $Z$ (7 mesh points), and $J$ (30 mesh points) were performed. In the case of the oblate ansatz (red dots, 36 points) this eventually results in a total amount of kernels of 979020 ($\frac{36\times37}{2}\times7\times7\times30$). 

The HWG equation~\eqref{eq:HWG} is eventually solved, to find the eingenvalues and eigenvectors (linear coefficients in the ansatz) of the Hamiltonian in the explored Hilbert subspace. This allows to provide the strength function of the investigated system. The monopole response in $^{28}$Si is displayed in Fig.~\ref{fig:HFB_PES} (right). The response in red corresponds to the oblate ansatz, which is interpreted as monopole vibrations on the oblate $^{28}$Si ground state. The response in yellow corresponds, instead, to the prolate ansatz, thus describing monopole vibrations on the prolate-shape isomer. The striking difference between the two responses showcases the flexibility of the PGCM, allowing to describe different kinds of physics according to the explored portion of the Hilbert space. In this specific case, the strong low-energy ($\sim$11~MeV) component in the prolate case also provides an excellent excellent example of the well-known coupling between quadrupole and monopole degrees of freedom in strongly deformed nuclei, see \cite{Peru08a,Porro24a,Porro24b} for further readings on this topic.

\subsubsection{The generator coordinate method and the random phase approximation}
Differently from (Q)RPA, the GCM is by construction able to capture many-body correlations beyond the harmonic hypothesis. 
Anharmonic effects may have a non-negligible impact on the determination of the nuclear incompressibility (\cite{Blaizot95a}), hence making the GCM a necessary tool in this respect. 
Indeed, it has been shown that the (Q)RPA wave function is found as the harmonic limit of the GCM, if one makes a quadratic approximation around the minimum of the energy manifold, provided that the basis states $\ket{\Phi(q)}$ span the entire one-body operator space (\cite{Jancovici64a,Brink68a,Brink68b,Federschmidt85a}).
The explicit mixing of basis states within the (P)GCM ansatz is thus capable of exploring the effects associated with the existence of several minima and anharmonicities which are not accessible, instead, to the (Q)RPA.
It is important to stress that every benefit entails a cost: either one explores the full operator space at the harmonic level, like in the (Q)RPA, or one can exactly treat anharmonic effects within the GCM, but just for a selected number of collective coordinates $q$.
Eventually, the success or not of (P)GCM calculations depends on the ability to select the most significant degrees of freedom of the system under exam.
\begin{figure}[h!]
    \centering
    \includegraphics[width=\linewidth]{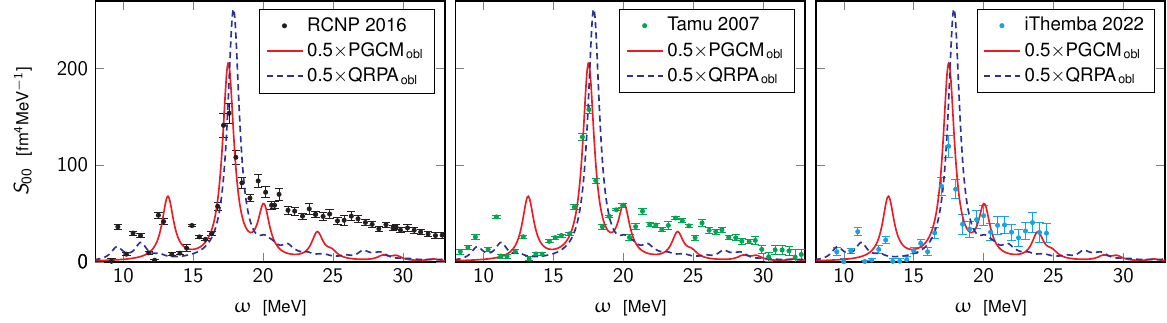}
    \caption{Monopole response in $^{28}$Si from PGCM and QRPA calculations employing the N3LO potential from~\cite{Huther19a}. The red curve is the results of the PGCM calculation employing the oblate ansatz (red dots in Fig.~\ref{fig:HFB_PES} (left)), while the dashed blue curve is the result of a QRPA calculation performed in the oblate HFB minimum. Experimental data are taken from~\cite{Peach16a,Youngblood07a,Bahini21a}.}
    \label{fig:Si28_exp}
\end{figure}

As a comparison, results from PGCM and QRPA monopole calculations, performed in a consistent setting, are displayed in Fig.~\ref{fig:Si28_exp} and compared to experimental data from three different campaigns (\cite{Youngblood07a,Peach16a,Bahini21a}). The QRPA calculations were performed based on the oblate deformed HFB minimum from Fig.~\ref{fig:HFB_PES} (left). One observes that QRPA calculations display a main resonance close in energy to the PGCM results, while the latter produce a richer fragmentation in the response over the entire energy domain.
%the response is overall much less fragmented over the entire energy domain.
When comparing to the different experimental datasets, PGCM calculations show a better agreement for the position of the main resonance ($\sim$17 MeV), as well as a better description of smaller  structures appearing both below and above the giant resonance. 
Eventually, while being less relevant for heavier systems, anharmonic effects seem to have an important role in the physics of monopole resonances in light- and medium-mass deformed nuclei like $^{28}$Si, for which PGCM calculations offer deeper insight than traditional QRPA calculations.

Finally, it should be stressed that the PGCM and QRPA differ as far as the treatment of symmetry restoration is concerned. While the symmetries of the Hamiltonian are explicitly restored within the PGCM, it is not so within standard (Q)RPA calculations relying on a symmetry-breaking HF(B) reference state.
%While the (Q)RPA partially contribute to restore some of the symmetries missing at the HF(B) level (\cite{Lane80a}), such restoration is not exact. 
Angular-momentum projection of deformed RPA calculations (\cite{Porro23a}) has shown that a non-negligible spurious coupling between rotational and vibrational degrees of freedom is present. 
A variationally exact treatment of such coupling has been only theoretically formulated so far (\cite{Federschmidt85a}), while no realistic implementation of this method exists yet.

\subsection{Self-consistent Green's functions theory}
\label{sec: green functions}

The self-consistent Green's functions (SCGF) approach, reviewed in Refs. \cite{Barbieri2004,Barbieri2017,Soma2020}, belongs to the class of many-body expansion methods and is rooted in the theory of Green's functions or propagators (\cite{Mattuck,Fetter,DickhofVanNeck}).
At variance with CC, the ground-state wave function is never parametrized explicitly in SCGF.
Rather, the key object is the one-body propagator $g(\omega)$, which is determined as the solution to the Dyson equations
\begin{align}
    \label{eq: dyson}
    g_{\alpha\beta}(\omega) =
    g_{\alpha\beta}^{(0)}(\omega) +
    \sum_{\gamma\delta}
    g_{\alpha\gamma}(\omega) \Sigma^{\star}_{\gamma\delta}(\omega) g_{\delta\beta}^{(0)}(\omega),
\end{align}
where Greek indices refer to single-particle~states. The function
$g^{(0)}(\omega)$ is a reference propagator representing a first approximation to $g(\omega)$, while $\Sigma^{\star}(\omega)$ is known as the irreducible self-energy (see below). 
For simplicity, we focus here on closed-subshell nuclei, where $g^{(0)}(\omega)$ is typically associated with a Hartree-Fock mean-field solution.
An extension to open-shell semi-magic nuclei, named Gorkov-SCGF, is discussed in, e.g., Refs.~\cite{Soma2011,Soma2020,Soma2020Chiral,Barbieri2022Gorkov}.

The one-body propagator can be interpreted as a transition amplitude for a nucleon propagating in the correlated nuclear medium, or as an energy-dependent one-body density.
Importantly, the propagator grants access to the total binding energy, as well as the expectation values of one-body operators (\cite{Soma2020}).
Additionally, it contains information on the neighboring isotopes with mass $A \pm 1$, in particular, the excitation energies (measured with respect to the $A$-nucleon ground state) and the transition amplitudes. This is evident in the spectral representation
\begin{align}
    \label{eq: lehmann repr gab}
    g_{\alpha\beta}(\omega) = \sum_n \frac{ \left( \mathcal{X}_\alpha^n\right)^* \mathcal{X}_\beta^n }{\hbar \omega - \epsilon_n^+ +i\eta} +
    \sum_k \frac{ \mathcal{Y}_\alpha^k ( \mathcal{Y}_\beta^k )^* }{ \hbar \omega - \epsilon_k^- - i \eta}\,,
\end{align}
where the one-nucleon addition and removal energies $\epsilon_n^+ = E_n^{A+1} - E_0^A$ and $\epsilon_k^- = E_0^A - E_k^{A-1}$ and the spectroscopic amplitudes $( \mathcal{X}_\alpha^n )^{*} = \mel{\Psi_0^A}{c_\alpha}{\Psi_n^{A+1}}$ and $\mathcal{Y}_\alpha^k = \mel{\Psi_k^{A-1}}{c_\alpha}{\Psi_0^A}$ have been introduced.
Here, $\ket{ \Psi_0^A }$ is the g.s. of the starting nucleus with $A$ nucleons, while $\ket{ \Psi_n^{A\pm1} }$  refer to the excited states of the $A \pm 1$ neighboring nuclei.
Notice that, within the CC framework, similar information can be obtained by performing particle-attached/particle-removed EOM-CC calculations on top of the ground state, see (\cite{Gour2006,Bartlett2007}).

The Dyson equation is formally exact. However, the self-energy must be necessarily approximated by retaining appropriate classes of Feynman diagrams.
In the SCGF formulation, $\Sigma^{\star}$ is expressed as a function of the dressed propagator $g(\omega)$ itself, which simultaneously determines and is determined by the self-energy, hence requiring to search for a self-consistent solution.
The accuracy of the SCGF predictions is tied to the quality of the self-energy ansatz.
Over the last decade, state-of-the-art nuclear physics computations have been achieved using the algebraic diagrammatic construction framework (\cite{Soma2011,Schirmer2018,Soma2020}).
This scheme has the attractive feature of providing a systematic hierarchy of approximations to the self-energy, and calculations based on the second-order and third-order truncations for open-shell and closed-subshell isotopes, respectively, compare well with coupled-cluster predictions at the triples level for both ground-state energies and densities (\cite{Soma:2013xha}).

The SCGF language is also suited for response functions (\cite{DickhofVanNeck,Raimondi2019}).
However, to access excited states of the $A$-particle nucleus, we must consider a specific two-body propagator, namely, the polarization introduced in Eq.~\eqref{eq: Polarization spectral repr}.
The polarisation propagator satisfies the so-called Bethe-Salpeter equation, which reads schematically as (\cite{Barbieri2003,Raimondi2019})
\begin{equation}
    \Pi(\omega) = 
    \Pi^{f}(\omega) +
    \Pi^{f}(\omega) K^{ ( \rm{ph} ) }(\omega) \Pi(\omega),
\end{equation}
where $\Pi^{f}(\omega)$ is the free polarization and $ K^{ ( \rm{ph} ) }$ is the particle-hole irreducible interactions, which plays for the polarization propagator a role similar to that of the self-energy for the one-body propagator.
The simplest approximation to $K^{ ( \rm{ph} ) }$ consists of using the bare interaction matrix elements.
If $\Pi^{f}(\omega)$ is built out of HF propagators, one obtains the usual RPA equations, Eq.~\eqref{eq: RPA matrix}.
An improvement over RPA is achieved if the reference polarization is constructed using correlated one-body propagators $g(\omega)$ from a preliminary SCGF computation, which leads implicitly to the inclusion of \NpartNhole{2} contributions to the polarization.
This scheme has been denoted as dressed RPA, and represents a hybrid approach, as the  ground state description is improved, while the interaction kernel is still approximated at first order (\cite{Raimondi2019}).
The effect of dressed RPA is to push excited states to higher energies (\cite{Barbieri2003,Raimondi2019}), while at the same time producing additional fragmentation in the response function.
SCGF calculations in nuclear physics have exploited this approximation to predict the dipole strength of closed-shell isotopes (\cite{RaimondiProceeding,Raimondi2019}). We discuss these results in Sec.~\ref{sec:benchmark_dipole}.
We mention in passing that an  algebraic diagrammatic construction hierarchy for the polarization propagator has been developed in chemistry (\cite{Schirmer2018}).

\subsection{No-core shell model and in-medium similarity renormalization group}
\label{sec: other methods}

Strength functions or sum rules thereof can also be obtained within the no-core shell model (NCSM) and the in-medium similarity normalization group (IMSRG) methods. 

As explained by \cite{Navratil:2009ut,Barrett2013_NCSM}, in the NCSM, the eigenstates of the Hamiltonian are expressed as
\begin{equation}
|\Psi_0\rangle = \sum_j \alpha_j | \Phi_j\rangle \,,
\end{equation}
where $|\Phi_j\rangle$ are Slater determinants constructed from harmonic-oscillator single-particle states. The expansion includes all many-body configurations of $A$ nucleons up to a maximum total excitation energy, defined by the truncation parameter $N_{max}$, which limits the total number of oscillator quanta above the lowest configuration. 
The coefficients $\alpha_j$ are obtained by diagonalizing the Hamiltonian matrix in this basis, which yields the ground state and low-lying excited states of the system. By systematically increasing $N_{max}$, the calculation converges toward the exact solution of the many-body Schrödinger equation for the chosen Hamiltonian. 

In practice, achieving convergence in the NCSM can be challenging, especially for heavier nuclei or for hard nuclear interactions. To facilitate convergence, one can soften the nuclear interaction using techniques such as the similarity renormalization group (SRG) as in \cite{Bogner2007_SRG}, which systematically decouples high- and low-momentum components. Furthermore, one can reduce the computational cost by selecting only the most relevant many-body basis states, as in the importance-truncated NCSM  introduced by \cite{Roth2007}, where the basis is truncated according to a perturbatively estimated contribution of each configuration to the target state. These approaches allow for accurate calculations while keeping the model space manageable.

To compute response functions within the NCSM, one can employ the Lanczos strength-function method initiated  by \cite{Whitehead1980}, see also \cite{Haxton2005}, which efficiently reconstructs the spectral distribution of a  transition operator without requiring full diagonalization of the Hamiltonian. Starting from the  correlated NCSM ground state $|\Psi_0\rangle$, one applies the relevant transition operator  $O$ to generate the pivot vector $ O|\Psi_0\rangle/\sqrt{\langle \Psi_0 |OO|\Psi_0\rangle}$. 
This vector is used to initialize a Lanczos iteration with the intrinsic Hamiltonian $H$. The Lanczos algorithm builds an orthonormal Krylov subspace 
%\[
%\mathcal{K}_n = \mathrm{span}\{\,|v_1\rangle,\, H|v_1\rangle,\, H^2|v_1\rangle,\,\ldots \, \},
%\]
within which the Hamiltonian is tridiagonalized, producing the recursive Lanczos coefficients 
$\{a_i,b_i\}$, see~\cite{Lanczos1950}. These coefficients define a continued–fraction representation of the Green's 
function and thus of the response function
\begin{equation}
\label{resp_gf}
R(\omega) = - \langle \Psi_0 |O^{\dagger}O|\Psi_0\rangle 
\frac{1}{\pi}\,\mathrm{Im}\,\langle \Psi_0 | O^{\dagger}
(\omega + E_0 - H + i\eta)^{-1}  O | \Psi_0\rangle \,,
\end{equation}
in the limit of $\eta \rightarrow 0$.
After a certain number of Lanczos iterations, which can go as far as 1000, see \cite{Stumpf2018}, the approximate eigenvalues and transition 
strengths converge.  The result is a discrete strength, due to the fact that one has discretized the continuum final states on a bound-basis. Such discretized strength shows a larger number of contributions as the number of Lanczos iterations and the model-space dimension are increased.

The discretized strength distribution can be smoothed out by taking a finite  $\eta$ to attempt a comparison to experiment, where $\eta$ is chosen to be similar to the experimental resolution,  e.g.  $\eta=1$ or 2 MeV. In such a case,
as shown in \cite{Stumpf2018}, the number of Lanczos steps needed to converge is significantly lower, e.g. about 50. 
The Lanczos strength function approach, hence, avoids the need for full diagonalization and makes the computation of electromagnetic response functions feasible and efficient in large NCSM 
spaces. 
The folded response function is essentially the LIT with $\Gamma=\eta$. 
The difference with respect to the LIT method is that the folded response is interpreted as a response function and not as an integral transform that needs to be inverted. We remark, however, that the folded response function in the Lanczos strength function approach depends on the chosen folding width. Furthermore, it often presents a residual dependence on the harmonic oscillator parameters and on the basis truncation, as shown in \cite{Stumpf2017,Stumpf2018}.  

The IMSRG method has been introduced in~(\cite{ChapterFossezHergert}).
%\textcolor{red}{Fossez and Hergert's contribution to this Encyclopedia}. 
However, a short description adapted to the targeted context is also given here. The idea of the IMSRG~(\cite{Tsukiyama10a,Hergert15a}) is to evolve the initial Hamiltonian normal-ordered to a reference state, $H(0)=H$, via a continuous series of unitary transformations $U(s)$ to decouple particle-hole excitations on top of the reference state,
\begin{equation}
    H(s)=U(s)H(0)U^\dagger(s)\, .
\end{equation}
The series of unitary transformations can be cast as a flow equation with the flow parameter $s$
\begin{equation}
    \label{eq: imsrg evolution}
    \frac{dH(s)}{ds}=[\eta(s),H(s)]\,,
\end{equation}
where $\eta(s)$ is the anti-Hermitian generator of the transformation
\begin{equation}
    \eta(s)=\frac{d\,U(s)}{ds}U^\dagger(s)=-\eta^\dagger(s)\,.
\end{equation}
Several strategies have been used to adapt this approach, initially tailored to the ground state, to the description and evaluation of nuclear excited states. The most popular choice is by far the so-called Valence-Space IMSRG (VS-IMSRG) explained in \cite{Tsukiyama12a,Bogner14a,Stroberg16a,Hergert16a,Stroberg19a,Miyagi20a}.
Within the VS-IMSRG, a frozen core and a valence space above this core are selected within the Hilbert space.
States outside the valence space are decoupled by the IMSRG flow Eq.~\eqref{eq: imsrg evolution}.
In this way, an effective Hamiltonian is produced in a systematically improvable way, which can then be diagonalized in the relatively small valence space using standard shell-model techniques. This method has proven extremely effective in providing an accurate description of low-lying states, but the restricted dimension of the valence space prevents its use to describe highly collective physics, like giant resonances or rotational states, even if current attempts to increase the dimension of the valence space have proven their effectiveness in describing rotational spectra of well-deformed systems, at the price of renouncing to exact diagonalization (see the work by~\cite{Cao25a}). 
A recent calculation of the Gamow-Teller response function of $^{78}\rm{Ni}$ has also been put forward in \cite{Li2025GamowTeller}.
Similarly to the CC case, the EOM method has also been explored within the IMSRG (see~\cite{Parzuchowski16a,Parzuchowski17a}), showing promising results in the description of electromagnetic observables from an ab initio standpoint. While this method would represent an optimal frame for the study of giant resonances, it has not been explored systematically so far. 
In addition, the IMSRG has recently been employed to evaluate the moments of the response function in doubly-closed shell nuclei throughout the nuclear chart (\cite{Porro2025,Bonaiti25Monopole}).
Sum rules can in fact be expressed as ground-state expectation values, which can be conveniently computed with IMSRG, without having to compute the excited-state spectrum.
Selected results of this work will be shown in the next section.

\section{Benchmark nuclei}
\label{sec: results}

In previous sections, several methods were introduced, highlighting  selected results for some of them, but no comparison between different techniques was performed. In the following, we show a critical comparison of results from different many-body methods, that employ the same nuclear interaction within a given nucleus. We discuss first the dipole response and then moments of the monopole response for the $^{16}$O and  of $^{40}$Ca nuclei.
We choose $^{16}$O and $^{40}$Ca as benchmark nuclei because they are among the few cases for which a direct comparison is possible. In addition, their doubly magic character  makes them suitable for both methods that include dynamical correlations and methods that focus on collective correlations.

\subsection{Isovector dipole response}
\label{sec:benchmark_dipole}

\begin{figure}[h!]
    \centering
    \includegraphics[width=0.45\linewidth]{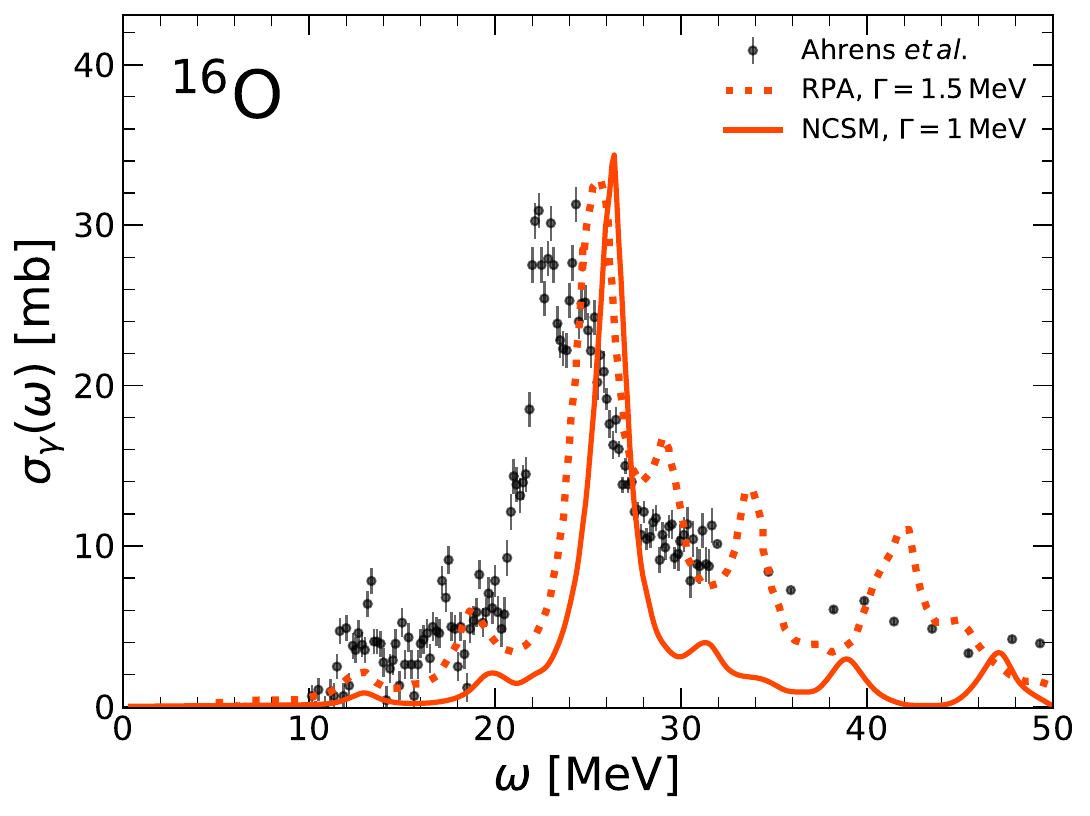}
    \includegraphics[width=0.45\linewidth]{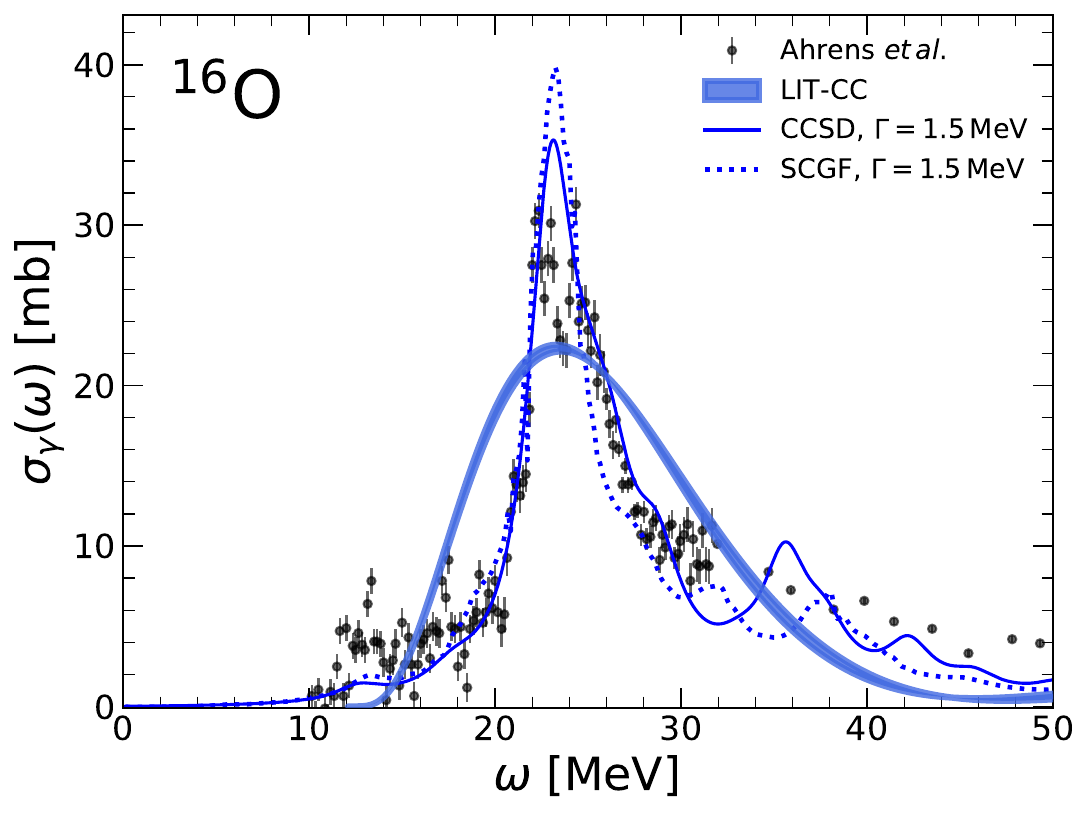}
    \includegraphics[width=0.45\linewidth]{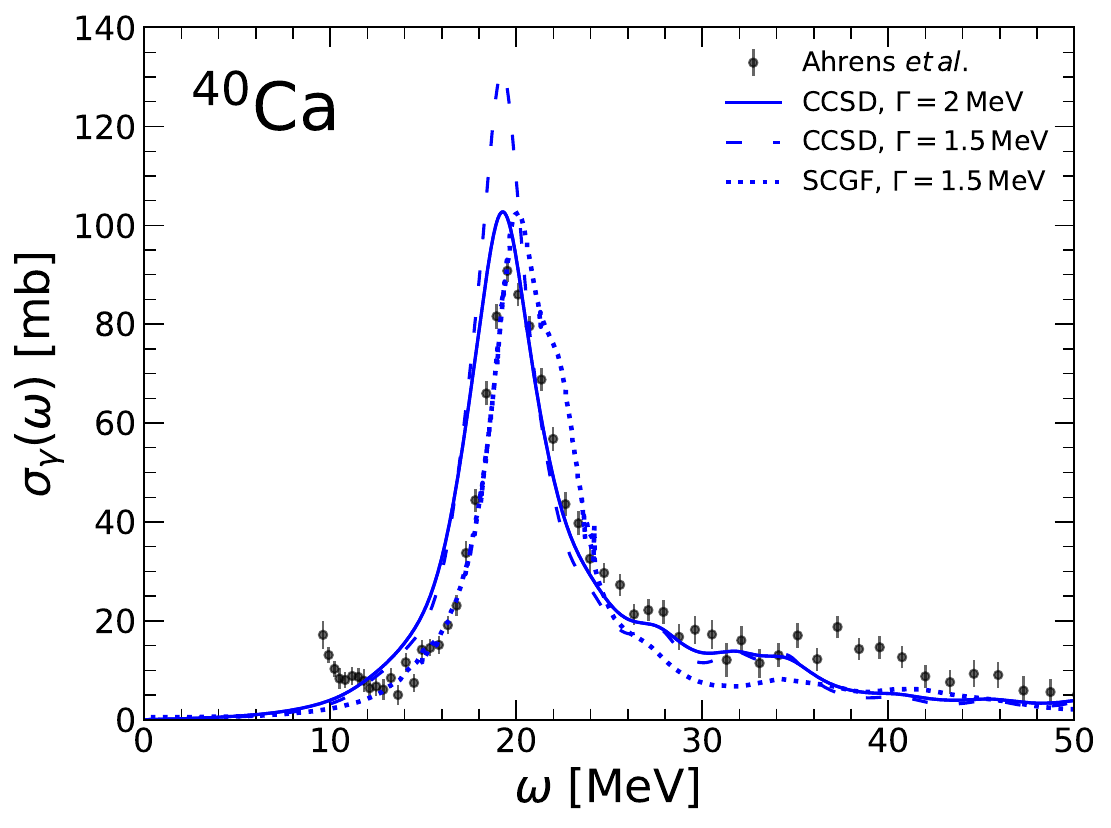}
    \caption{
    Photoabsorption cross sections of $^{16}$O and $^{40}$Ca as a function of the excitation energy.
    Experimental data are taken from \cite{AhrensO16,Ahrens1985}.
    The top left panel shows discretized responses folded with Lorentzians of width $\Gamma$ (reported in the legend) computed with RPA (\cite{BeaujeaultTaudiere2022}) and NCSM (\cite{Stumpf2017}) using an SRG-evolved interaction based on $\rm{ NNLO_{sat} }$.
    The top right and bottom panels both report folded discretized responses calculated from the CCSD (\cite{Miorelli2016}) and SCGF (\cite{Raimondi2019}) methods using the bare $\rm{ NNLO_{sat} }$ potential.
    For $^{16}$O, the thicker curve (LIT-CC) (from \cite{Miorelli2016}) represents the cross section obtained by inverting the LIT, with an estimate of the uncertainties associated with the inversion.
    RPA uses a harmonic-oscillator basis with  $N_{max}=12$ and oscillator frequency $\hbar\omega = 20\,\rm{MeV}$
    NCSM uses $N_{max}=9$ and $\hbar\omega = 20\,\rm{MeV}$.
    All the SCGF calculations are performed in a basis with $N_{max}=14$ and $\hbar\omega = 22\,\rm{MeV}$.
    CCSD and LIT-CC for $^{16}$O employ $N_{max}=14$ and $\hbar\omega = 22\,\rm{MeV}$, while $\hbar\omega = 14\,\rm{MeV}$ is used for $^{40}$Ca.
    }
    \label{fig:photoabs O16 two panels}
\end{figure}

As previously discussed, photoabsorption cross sections $\sigma_\gamma$ are related to the isovector electric dipole response functions by Eq.~\eqref{cs_siegert}.
Ab initio predictions for $\sigma_\gamma(\omega)$ are shown in Fig.~\ref{fig:photoabs O16 two panels} for $^{16}$O and $^{40}$Ca,  compared to the experimental data from \cite{AhrensO16,Ahrens1985}.
Let us first comment on the top left panel, where RPA and NCSM calculations are reported for $^{16}$O. These are taken from (\cite{BeaujeaultTaudiere2022}) and (\cite{Stumpf2017}), respectively, and are computed starting from the same interaction,
derived from the $\rm{NNLO_{sat}}$ potential (\cite{NNLOsat}) and softened by applying an SRG transformation (\cite{Roth2011,Hergert:2015awm}).
The discrete response functions have been folded with a Lorentzian of width $\Gamma = 1.5 \,\rm{MeV}$ and 1 MeV for RPA and NCSM, respectively.
Both RPA and NCSM feature a dominant peak  around the energy range of the experimentally observed GDR. The height of the peak is consistent with the measured cross section, although both calculations tend to overestimate the centroid energy of the resonance by a few MeV, with the NCSM result being shifted to slightly higher energies.
In the high-energy tail, the RPA cross section shows some relatively large bumps. In contrast, NCSM predictions lack most of the high-energy strength at $\omega \ge 30\,\rm{MeV}$.

In the top-right panel,  results for $^{16}$O obtained using the bare $\rm{NNLO_{sat}}$ interaction are displayed.
The curve labeled LIT-CC from (\cite{Miorelli2016}) corresponds, similarly to  Fig.~\ref{fig:LIT-CC-results}, to the photoabsorption cross section extracted by inverting the LIT computed at the CCSD level, but this time with the $\rm{NNLO_{sat}}$ interaction, which includes three-nucleon forces.
The shaded band represents an uncertainty estimate arising from the inversion procedure, reflecting the spread in the response functions obtained when using different values of the width parameter $\Gamma$ between 10 and 20 MeV. 
The centroid of the GDR is well reproduced. Moreover, the corresponding electric dipole polarizability,
\begin{align}
    \alpha_D = \frac{\hbar c}{2\pi^2} \int_{0}^{\infty}~d\omega \frac{ \sigma_{\gamma}(\omega)}{ \omega^2 }
\end{align}
evaluated in \cite{Miorelli2016}, agrees well with the experimental value $\alpha_D = 0.58\,(1)\,\rm{fm}^{3}$ from \cite{AhrensO16}. The inclusion of triples excitations in the ground state would reduce $\alpha_D$ by a small amount (\cite{Miorelli2018}).
The main shortcoming of the LIT-CC result is that the GDR appears too broad, which also results in a lower peak height compared to the experimental data. However,  improvements in the inversion procedure and a better quantification of its associated uncertainties could possibly enhance the results.

The folded discretized responses from CCSD and SCGF are also shown in Fig.~\ref{fig:photoabs O16 two panels}, with the corresponding Lorentzian widths reported in the legend.
If the width of the LIT is fixed to approximately reproduce the observed width of the resonance ($\Gamma = 1.5\,\rm{MeV}$), an improved description of experimental data is obtained. The corresponding result provides a reasonable description not only of the GDR centroid but also of its peak height. In addition, it predicts dipole strength at low energies (between 10 and 20 MeV) as well as in the high-energy tail of the resonance, extending up to about 30 MeV.
It is important to emphasize that the centroid position and the integral sum rules, such as $\alpha_D$, do not depend on the choice of the width parameter (provided it is not too large); they represent genuine predictions of the coupled-cluster calculation. By contrast, the width and peak height of the response function are strongly correlated and sensitive to the adopted value of $\Gamma$.
From this perspective, performing an inversion of the LIT is, in principle, preferable, since it mitigates the residual arbitrariness associated with the choice of the folding parameter.

The SCGF calculation reported in Ref.~(\cite{Raimondi2019}) also shows reasonable agreement with the experimental data in the energy region of the GDR. In particular, the centroid energy is well reproduced. The use of a correlated propagator in constructing the RPA matrices is essential, leading to a substantial improvement over a standard RPA calculation based on a Hartree–Fock reference (\cite{Raimondi2019,BeaujeaultTaudiere2022}).
The width of the GDR is reproduced by construction through the Lorentzian folding with $\Gamma = 1.5\,\rm{MeV}$, and for this choice of the smearing parameter the overall magnitude of the cross section is also well described. However, as already observed for CCSD, the dressed RPA results exhibit a strong sensitivity to the value of $\Gamma$, which somewhat limits the predictive power of the present many-body truncation.
Some deficiencies of the current SCGF approximation become apparent at excitation energies above the GDR region, where a reduction of strength is observed compared to the experimental data. This missing strength is attributed to couplings with more complex particle–hole configurations that lie beyond the RPA framework and would require higher-order many-body treatments.
The SCGF approach also yields a slightly smaller electric dipole polarizability, $\alpha_D \approx 0.50,\mathrm{fm}^3$, compared to both the CC results and the experimental value. This underestimation is likely related to a deficit of low-energy strength, to which $\alpha_D$ is particularly sensitive~(\cite{Miorelli2016}).
In summary, both the CCSD and SCGF calculations provide a reasonable, though not fully quantitative, description of the data and successfully capture the main physical features of the GDR.

It is worth emphasizing that calculations based on the bare $\rm{NNLO_{sat}}$ interaction yield overall more satisfactory results than the RPA and NCSM calculations shown in the top-left panel, particularly with regard to the position of the GDR peak. We are led to attribute most of the observed discrepancies to the underlying Hamiltonian rather than to the specific ab initio many-body method employed, as we now discuss.
In this context, it is important to recall that the SRG evolution induces three- and many-nucleon contributions in the transformed Hamiltonians, which are discarded in practical applications within many-body frameworks. As a consequence, the SRG-evolved potential is in effect a different interaction, and some of the properties of the original model may be altered or partially lost in the evolution and truncation procedure.
The bare $\rm{NNLO_{sat}}$ potential, on the other hand, is known to provide a good description of bulk nuclear observables—such as binding energies, charge radii, and form factors—up to medium-mass nuclei (see, e.g., Refs.~\cite{Payne:2019wvy,Soma2020Chiral,PbAbInitio,Tsaran:2025qfh}). Given the established correlation between the electric dipole response and the nuclear density distribution~(\cite{Piekarewicz2006,Raimondi2019}), one may therefore expect this interaction to capture the gross features of the GDR in $^{16}\mathrm{O}$ and other closed-shell systems.
This expectation appears to be confirmed by both the CCSD and SCGF results, which reproduce the main characteristics of the GDR with reasonable accuracy.

Finally, we turn to the photoabsorption cross section of $^{40}\mathrm{Ca}$, shown in the bottom panel of Fig.~\ref{fig:photoabs O16 two panels}. Cross sections from folded discretized responses obtained within the CCSD and SCGF frameworks using the bare $\rm{NNLO_{sat}}$ interaction are displayed.
For the CCSD results, we consider two choices of the folding width. While the location of the peak is unaffected, the peak height and the shape of the GDR are affected by $\Gamma$. 
With $\Gamma=1.5$ MeV, the height is overestimated, while for $\Gamma=2$ MeV the agreement with the experiment is considerably better.
The SCGF calculation, where the chosen width is 1.5 MeV, also provides a reasonable  overall description of the cross section. The peak is slightly offset by about 2 MeV from the CCSD prediction. This implies that the low-energy side of the resonance is better described than in CCSD, while the high-energy side is slightly further away from the experimental data.
As already observed in the case of $^{16}\mathrm{O}$, the SCGF calculation exhibits a reduction of strength in the high-energy tail of the spectrum, indicating limitations of the present dressed RPA approximation.
The LIT-CC result, on the other hand, shows better agreement with the experimental data up to excitation energies of about 35 MeV.
Overall, the situation for $^{40}\mathrm{Ca}$ closely parallels that discussed for $^{16}\mathrm{O}$: both CCSD and SCGF capture the main characteristics of the dipole response, while quantitative differences remain in the detailed distribution of strength.
Most importantly, these calculations show that collective modes emerge from first principles from the underlying nuclear interaction.

\subsection{Isoscalar monopole sum rules}
\label{sec: Monopole response}
Recently, different ab initio techniques have been applied to the computation of the isoscalar monopole sum rules in \cite{Bonaiti25Monopole}.
Results from RPA, CC and IMSRG calculations  were compared in doubly-closed-shell $N=Z$ nuclei in a consistent setting. 
The moments of the monopole response were then used to extrapolate the incompressibility of symmetric nuclear matter (\cite{Garg18a}).
CC sum rules have been determined from calculations at the CCSD level using Eq.\eqref{eq: moments def}. 
In IMSRG, these have been obtained as ground-state expectation values, see Ref.~(\cite{Porro2025}) and Sec.~\ref{sec: other methods}.
\begin{figure}[h!]
    \centering
    \includegraphics[width=0.9\linewidth]{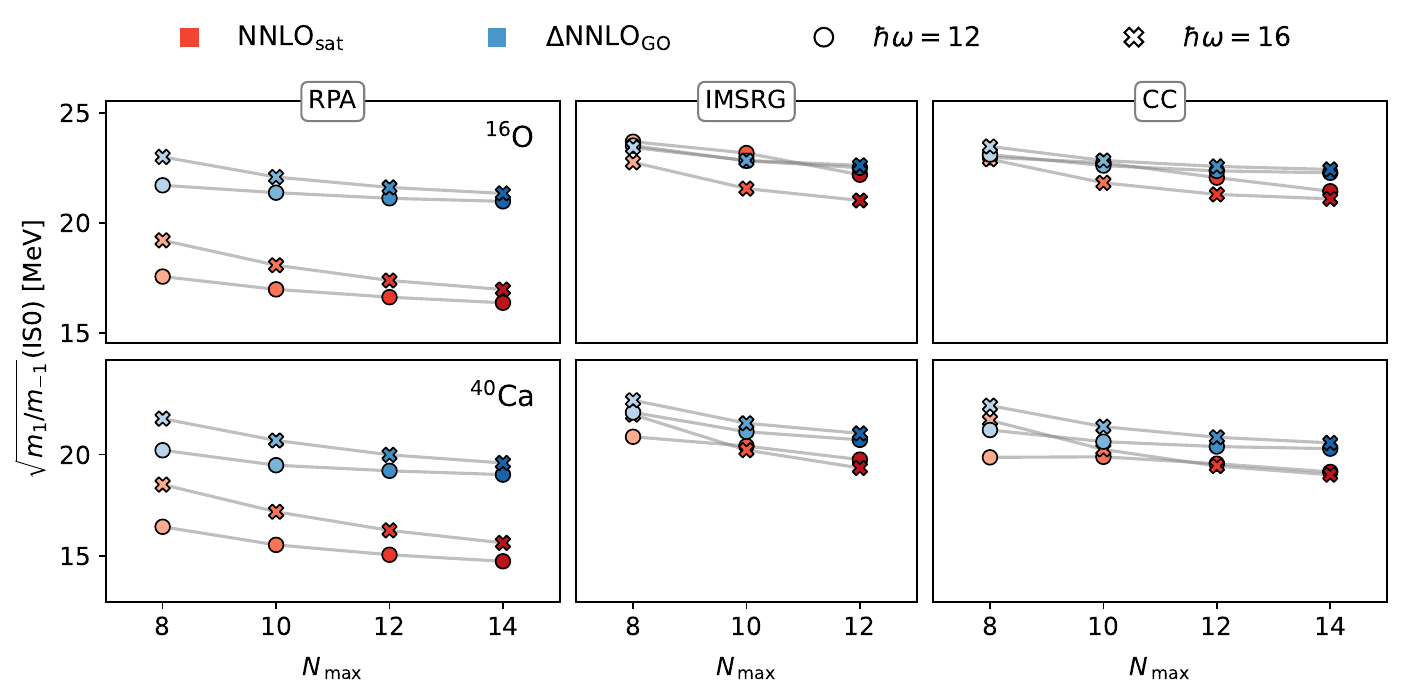}
    \caption{Average energy of the isoscalar monopole response in $^{16}$O and $^{40}$Ca, obtained with (from left to right) RPA, IMSRG, and coupled-cluster calculations. The horizontal axis displays different sizes $N_{max}$ of the harmonic oscillator basis employed in the calculations. 
    Results for two different frequencies $\hbar\omega = 12\,\rm{MeV}$ (16 MeV) are shown as circles (crosses).
    The figure has been adapted from~\cite{Bonaiti25Monopole}.}
    \label{fig:monopole_ave}
\end{figure}

The average energy, defined as the ratio
\begin{equation}
    \bar{E}\equiv\sqrt{\frac{m_1}{m_{-1}}}\,,
\end{equation}
is shown in Fig.~\ref{fig:monopole_ave} for three different many-body approaches and for two chiral Hamiltonians, namely $\rm{NNLO_{sat}}$~(\cite{NNLOsat}) and $\rm{\Delta NNLO_{GO}}(394)$~(\cite{Jiang20a}). The convergence pattern with respect to the model-space size is examined by displaying $\bar{E}$ as a function of $N_{max}$. In addition, results are reported for two different harmonic-oscillator frequencies $\hbar\omega$, allowing one to assess the residual dependence on the underlying single-particle basis.
The results appear substantially converged with respect to $N_{max}$, and the residual dependence on $\hbar\omega$ is often negligible, especially for IMSRG and CC.
A very good agreement between IMSRG and CC is observed for both interactions and for both nuclei considered. Moreover, as discussed in \cite{Bonaiti25Monopole}, the predicted average energies are in fair agreement with the experimental values (not shown in the figure).
The simpler RPA results display a more interaction-dependent behavior. When $\rm{\Delta NNLO_{GO}}$ is employed, RPA yields values that are in approximate agreement with IMSRG and CC, although they still tend to underestimate the average energies slightly. In contrast, significantly larger deviations are found for $\rm{NNLO_{sat}}$. This difference can be traced back to the nature of the two Hamiltonians.
$\rm{\Delta NNLO_{GO}}$ is a comparatively soft interaction, characterized by a lower cutoff (394 MeV versus 450 MeV for $\rm{NNLO_{sat}}$), which enables a faster convergence with respect to both the model-space dimension and the many-body approximation.
As a consequence, the dynamical correlations induced by $\rm{\Delta NNLO_{GO}}$ are less pronounced, and mean-field-based approaches such as RPA already provide a reasonable description of the excitation spectrum. This explains the relatively good agreement with CC and IMSRG, which explicitly incorporate \NpartNhole{2} and higher-order correlations in the many-body wave function. For $\rm{NNLO_{sat}}$, however, such correlations play a much more significant role, and they increase the average energy from about 15 MeV in RPA to 20 MeV or higher in CC and IMSRG calculations. These more sophisticated methods thus deliver a more realistic description of the GMR centroid, which is expected to dominate the value of $\bar{E}$.
The excellent agreement between CC and IMSRG results, largely independent of the chosen interaction, provides an important consistency check among different state-of-the-art many-body frameworks in the study of nuclear collective excitations. In Ref.~\cite{Bonaiti25Monopole}, the calculated average monopole energies were further employed to extract the incompressibility of finite nuclei. The extrapolation to infinite nuclear matter revealed a consistency between IMSRG and CC predictions. 
While the resulting incompressibility values are lower than those obtained in nuclear-matter calculations with the same interactions, they remain consistent with phenomenological constraints.

\section{Future Perspectives}

In this chapter, we have reviewed recent progress in the ab initio description of nuclear response functions, with particular emphasis on electromagnetic observables and collective excitations. The last decade has witnessed substantial advances driven by the development of chiral effective field theory interactions, increasingly sophisticated many-body methods, and rapidly growing computational resources. These achievements demonstrate that a quantitative and, in some cases, predictive description of nuclear response is now within reach. At the same time, several conceptual and practical challenges remain before a fully systematic framework can be established.

A first major limitation of present ab initio response calculations concerns their domain of applicability. Approaches that incorporate dynamical correlations in a controlled manner, such as self-consistent Green’s function theory and Lorentz integral transform coupled-cluster theory, have so far been applied predominantly to closed-(sub)shell nuclei. While recent extensions to open-shell isotopes via particle-attached and particle-removed coupled-cluster formulations represent an important step forward~(\cite{Bonaiti2024,Marino2025}), truly open-shell and deformed systems remain largely out of reach in fully dynamical treatments of the response. In contrast, multi-reference and symmetry-breaking approaches such as the projected generator coordinate method are, in principle, applicable to nuclei across the nuclear chart, including open-shell and deformed systems. However, in their present ab initio implementations with chiral Hamiltonians, PGCM-based calculations primarily capture static correlations, with dynamical correlations treated only approximately.

This complementarity highlights a central tension in current ab initio response theory: methods that include dynamical correlations with high fidelity are, for now, limited in terms of the nuclear systems they can address, while methods that can access the whole nuclear chart (or most of it) rely on reference states that lack a systematic inclusion of dynamical many-body correlations. Recent developments, such as perturbative corrections on top of PGCM reference states, provide an important proof of principle that dynamical correlations can be added on top of symmetry-breaking and symmetry-restored reference states, and that these corrections may partially cancel for low-lying collective excitations, see \cite{Frosini21a,Frosini21b,Frosini21c,Duguet22a}. Whether similar conclusions hold for giant resonances and high-energy collective response remains an open and important question.

From a broader perspective, further progress requires both extending the range of nuclei accessible to fully dynamical approaches and improving the treatment of correlations in multi-reference frameworks. Symmetry-breaking formulations of self-consistent Green’s function theory [Gorkov-SCGF (\cite{Soma2020})] and coupled-cluster theory [Bogoliubov CC (\cite{Tichai2024})] already offer promising routes to include pairing correlations and access open-shell nuclei, while recent extensions toward deformed reference states indicate that genuinely non-spherical systems may become tractable in the near future~(\cite{Hagen2022}). Ultimately, however, the restoration of broken symmetries—most notably angular momentum—will be crucial for a controlled description of resonances and electromagnetic response (\cite{Porro23a,Chen25a}).

Another challenge is the need for systematic and quantitative uncertainty estimates (\cite{PbAbInitio,Ekstrom2023}). This includes uncertainties associated with truncations of the many-body expansions, model-space limitations, and the input interactions derived from chiral effective field theory. Addressing these issues is essential if ab initio response calculations are to become predictive in a robust sense and reliably applicable to regions of the nuclear chart where experimental information is scarce or unavailable.

Finally, besides the $^{22}$O case of Fig.~\ref{fig:LIT-CC-results}, most applications discussed in this chapter concern stable nuclei. Extending ab initio response theory to exotic, neutron-rich systems represents a particularly compelling and largely unexplored frontier (\cite{Aumann:2024unk,Brown_2025}). Key open problems include the evolution of dipole strength with neutron number, the possible emergence of low-lying collective modes~(\cite{Paar:2007bk,Bracco2019,Lanza2023}), and the electromagnetic response of halo nuclei~(\cite{Aumann_2013}). In such weakly bound systems, the interplay between deformation, pairing, continuum coupling, and theoretical uncertainties becomes even more pronounced, further motivating the development of unified frameworks.

In the long term, a central goal of the field is to merge the complementary strengths of present-day approaches: the ability of symmetry-breaking and symmetry-restoration methods to describe deformation and collective motion across the nuclear chart, and the capability of approaches such as SCGF and LIT-CC to incorporate dynamical correlations in a controlled, systematically improvable manner. Achieving such a synthesis would mark a decisive step toward a truly universal ab initio theory of nuclear response, capable of providing quantitative predictions for electromagnetic and electroweak observables from light nuclei to heavy, deformed, and weakly bound systems.

\section*{Acknowledgements}
We thank Gianluca Col\`{o}, Weiguang Jiang, and Peter von Neumann-Cosel for useful feedback.

\bibliographystyle{Harvard}
\bibliography{bibliography}

\end{document}